\documentclass[12pt,preprint]{aastex}

\usepackage{graphicx}
\usepackage{amsmath}
\usepackage{amssymb}

\begin{document}

\title{An Updated Ultraviolet Catalog of \textit{GALEX} Nearby Galaxies}

\author{
   Yu Bai, 
   Hu Zou,
   JiFeng Liu,
   Song Wang
}
\affil{Key Laboratory of Optical Astronomy, National Astronomical Observatories, Chinese Academy of Sciences,
       20A Datun Road, Chaoyang Distict, 100012 Beijing, China; ybai@nao.cas.cn,
       zouhu@nao.cas.cn, jfliu@nao.cas.cn, songw@nao.cas.cn}

\begin{abstract}
The ultraviolet catalog of nearby galaxies made by \citet{Gil07} presents the integrated
photometry and surface brightness profiles for 1034 nearby galaxies observed by
\textit{Galaxy Evolution Explorer} (\textit{GALEX}). We provide an updated catalog of 4138
nearby galaxies based on the latest Genral Release (GR6/GR7) of \textit{GALEX}. These galaxies
are selected from HyperLeda with apparent diameter larger than 1{\arcmin}. From the
surface brightness profiles accurately measured with the deep NUV and FUV images, we
have calculated asymptotic magnitudes, aperture (D25) magnitudes, colors, structural
parameters (effective radii and concentration indices), luminosities, and effective
surface brightness. Archival optical and infrared photometry from HyperLeda, 2MASS,
and IRAS are also integrated into the catalog. Our parameter measurements and
some analyses are consistent with those of \citet{Gil07}. The (FUV $- K$) color
provides a good criterion to distinguish early and late-type galaxies,
which can be improved further with the concentration indices. The IRX-$\beta$
relation is reformulated with our UV-selected nearby galaxies.
\end{abstract}

\maketitle

\section{INTRODUCTION}
\label{intro}
Nearby galaxies (z $\la$ 0.1) are characterized by large angular scale and high apparent
brightness, which can be studied in more details and with higher accuracy than high-redshift
galaxies. For nearby galaxies, the ultraviolet (UV) imaging provides unique information. Massive,
young stars emit strong UV energy, which dominates the integrated UV light of star-forming
galaxies. Thus, the UV flux is widely used as an excellent and accurate measurement of
the current star formation rate (SFR; \citealt{Wilkins12,Kennicutt98,Lanz13}). In addition,
the interstellar dust can absorb the UV light and then re-emit at far-infrared wavelength.
The comparison between infrared and UV emission can effectively trace the dust attenuation
in galaxies. Radiative transfer models suggest that the ratio of far-infrared to UV
luminosity is a reliable estimator of the dust attenuation \citep{Buat96,Witt00,Panuzzo03},
depending weakly on the geometry of stars and dust, the extinction law, and the stellar
population.

The UV galaxy samples in the local universe are important for understanding the
evolution of galaxies with cosmic time \citep{Bouwens12,Reddy12,Ellis13,Martin05}. There
have been considerable attempts to explore the SFR, morphology, dust attenuation, and their
evolutions by constructing UV-selected local galaxies
\citep{Sullivan04,Calzetti94,Buat02,Kuchinski00,Marcum01,Lauger05}.

Based on \textit{GALEX} Public Release 2 and 3 (GR2/GR3), \citet{Gil07} (hereafter
\textit{GALEX} Atlas) presented a UV
catalog of 1034 nearby galaxies selected from the Third Reference Catalog of Bright
Galaxies (RC3; \citealt{de91}). They selected those galaxies whose optical diameters
of the $\mu_B = 25$ mag arcsec$^{-2}$ isophote (D25) are larger than 1{\arcmin}. 
The FUV and NUV images were mainly taken
from \textit{GALEX} Nearby Galaxies Survey (NGS). From the surface brightness profiles,
they obtained asymptotic magnitudes, colors, and concentration indices. In
combination with archival optical and infrared data, they analyzed the color-magnitude and
color-color diagrams, relations between colors and morphologies, IRX$-\beta$ relation,
dust attenuation, and structural properties for different types of galaxies.

In February 2013, \textit{GALEX} GR7 data was available. It includes more than
45000 images, almost three times larger than GR2/GR3. With deeper observations (up to
250 kilo-second), it is now possible to create a relative complete UV sample of nearby
galaxies. In this work, we provide an updated UV catalog of nearby galaxies with
deepest FUV and NUV images from GR6/GR7, archival optical and infrared photometry, and
make some similar analysis to \citet{Gil07}.

In Section~\ref{data}, we present our sample selection of nearby galaxies. In Section~\ref{sbm}, the method
of measuring the surface brightness profile is described in detail. The catalog content including the
UV parameters and other archival data are described in Section \ref{cc}. The parameter comparison
of our catalog with \textit{GALEX} Altas is shown in Section \ref{compga}. Section \ref{aa} presents
some analyses and application examples of our catalog. The summary is in Section \ref{sum}.
All magnitudes presented in the paper are corrected for the Galactic extinction using
the Galactic Reddening Map of \citet{Schlegel98}. We use the reddening law of  \citet{Cardelli89} to
convert $E(B-V)$ to the UV extinction:  $A_{FUV}=7.9E(B-V)$ and $A_{NUV}=8.0E(B-V)$.

\section{Sample Selection}
\label{data}
\begin{figure}
   \centering
   \includegraphics[width=8cm]{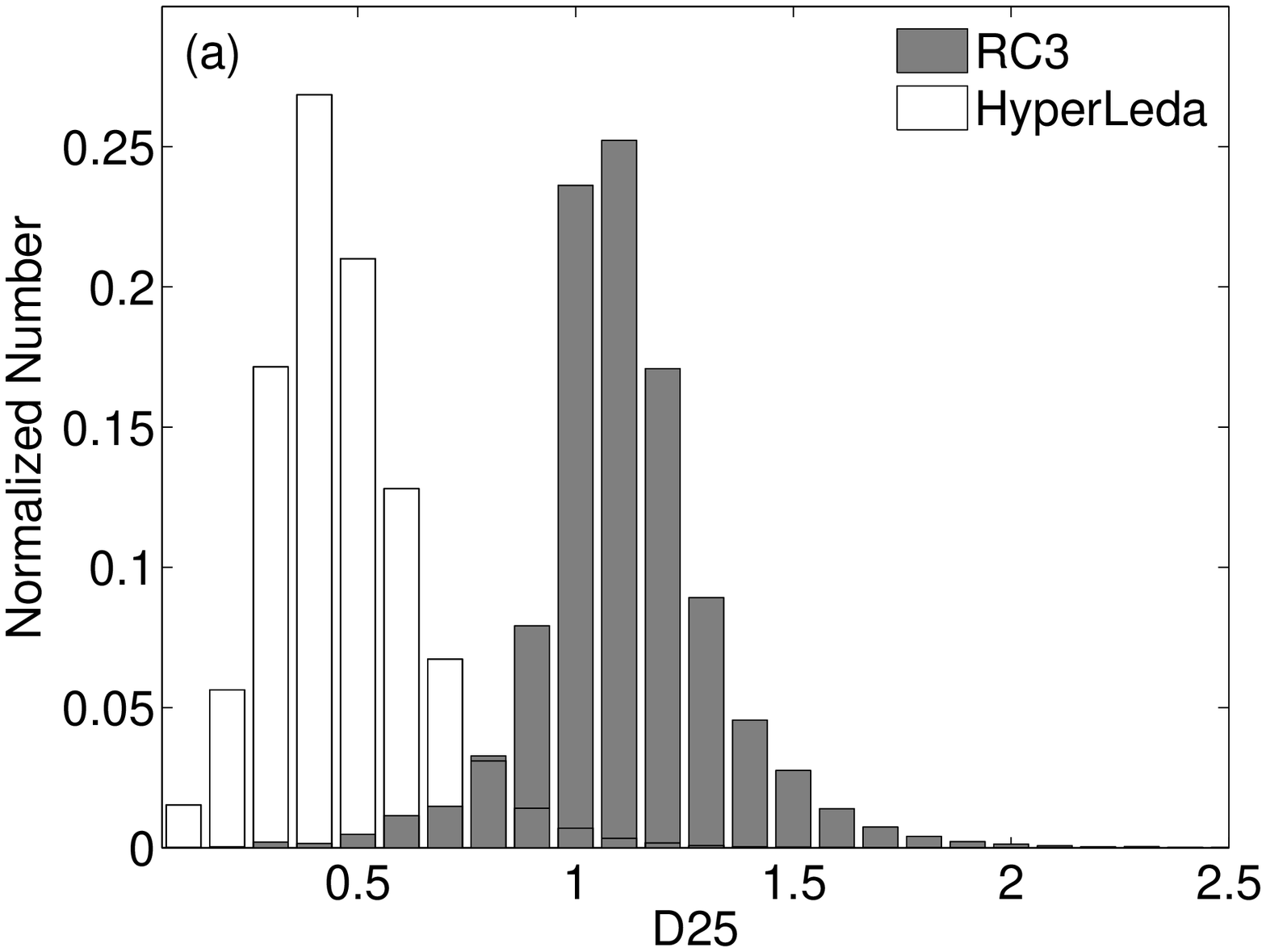}
   \includegraphics[width=8cm]{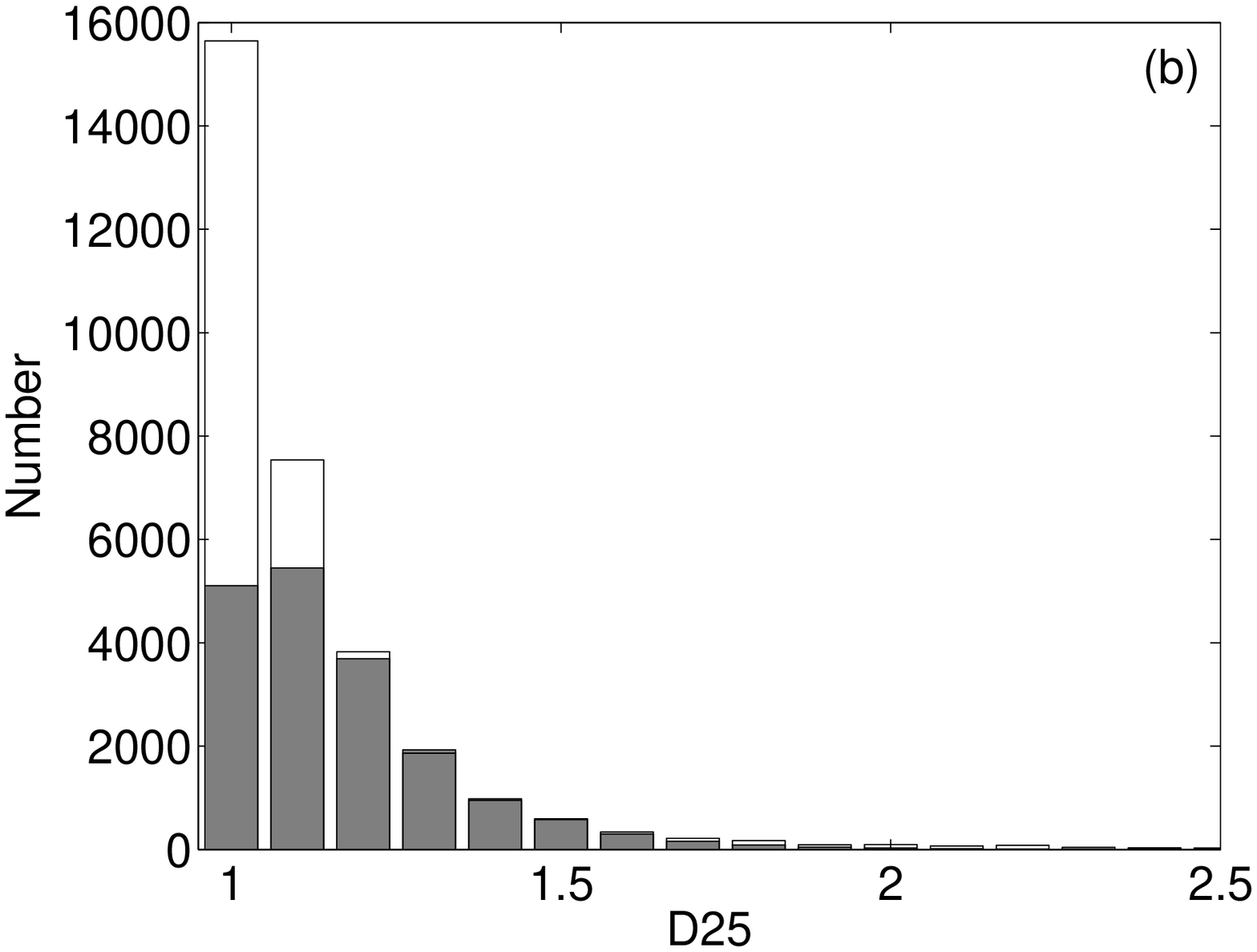}
   \caption{Comparison of D25 diameters in HyperLeada and RC3. (a) Normalized D25 distribution
	   of major-axis diameters. (b) D25 distribution of major-axis diameters larger than 1{\arcmin}.
   \label{fig1}}
\end{figure}

The nearby galaxies in this paper are extracted from HyperLeda \footnote{\url{http://leda.univ-lyon1.fr, revised in 2013 July 1}}
\citep{Paturel03,Makarov14}. HyperLeda is a catalog of
extragalactic sources, which gives accurate coordinates of objects (typical accuracy
better than 2$\arcsec$), morphological parameters (e.g. size, axis ratio, position
angle, morphogical type, etc.), and astrophysical parameters (optical magnitudes and colors,
surface brightness, distance, kinematic quantities, etc.). It includes 3.7 million galaxies,
much more than RC3 as used by \citet{Gil07}.
The catalog provides Doppler distances for 20314 galaxies and redshift-independent distances
for 2013 galaxies.
Figure \ref{fig1}a shows the distributions of major-axis diameters in both HyperLeda
and RC3 catalogs, while Figure \ref{fig1}b presents the same distributions but
with D25 $>1 \arcmin$. The majority of HyperLeda galaxies have diameters less than 1{\arcmin},
and over 99\% RC3 galaxies are included in HyperLeda.  The HyperLeda galaxies with diameters
above 1{\arcmin} are more than those of RC3 by a factor of 3.

We adopt the same criterion of optical diameters as \citet{Gil07} used, that is, D25 is larger than
1$\arcmin$. There are 22948 galaxies satisfying this constraint. We retrieve FUV and NUV images
for these galaxies from \textit{GALEX} GR6/GR7 to obtain their intensity maps. Three imaging
surveys are involved: Medium Imaging Survey (MIS), Nearby Galaxy Survey (NGS), and Deep Imaging
Survey (DIS). We require the galaxies (1.5 $\times$ D25) to be fully covered by
the \textit{GALEX} field of view (FoV; 1.2{\arcdeg} in diameter). Those galaxies with low
signal-to-noise ratio or contaminated by surrounding galaxies and bright stars are excluded.
The optical diameter D25 of M31 is about 3.0{\arcdeg}, far beyond the \textit{GALEX} FoV.
We obtain 26 tiles of M31 in FUV and NUV from NGS and DIS, which are stacked to
form very deep mosaics by Montage\footnote{\url{http://montage.ipac.caltech.edu/}}.
The final sample contains 4138 galaxies, 2321 of which have both FUV and NUV observations.
There are 14 and 1803 galaxies having only FUV and NUV observations, respectively.

We present their basic parameters given by HypeLeda in Table 1, including the
positions, sizes, morphological types and distances.
For galaxies whose distances are not available, we get their distances from NED (the
NASA/IPAC Extragalactic Database). There are 4047 galaxies with either luminosity distances
or redshift-dependent distances.

\begin{figure}
\centering
\includegraphics[width=45mm]{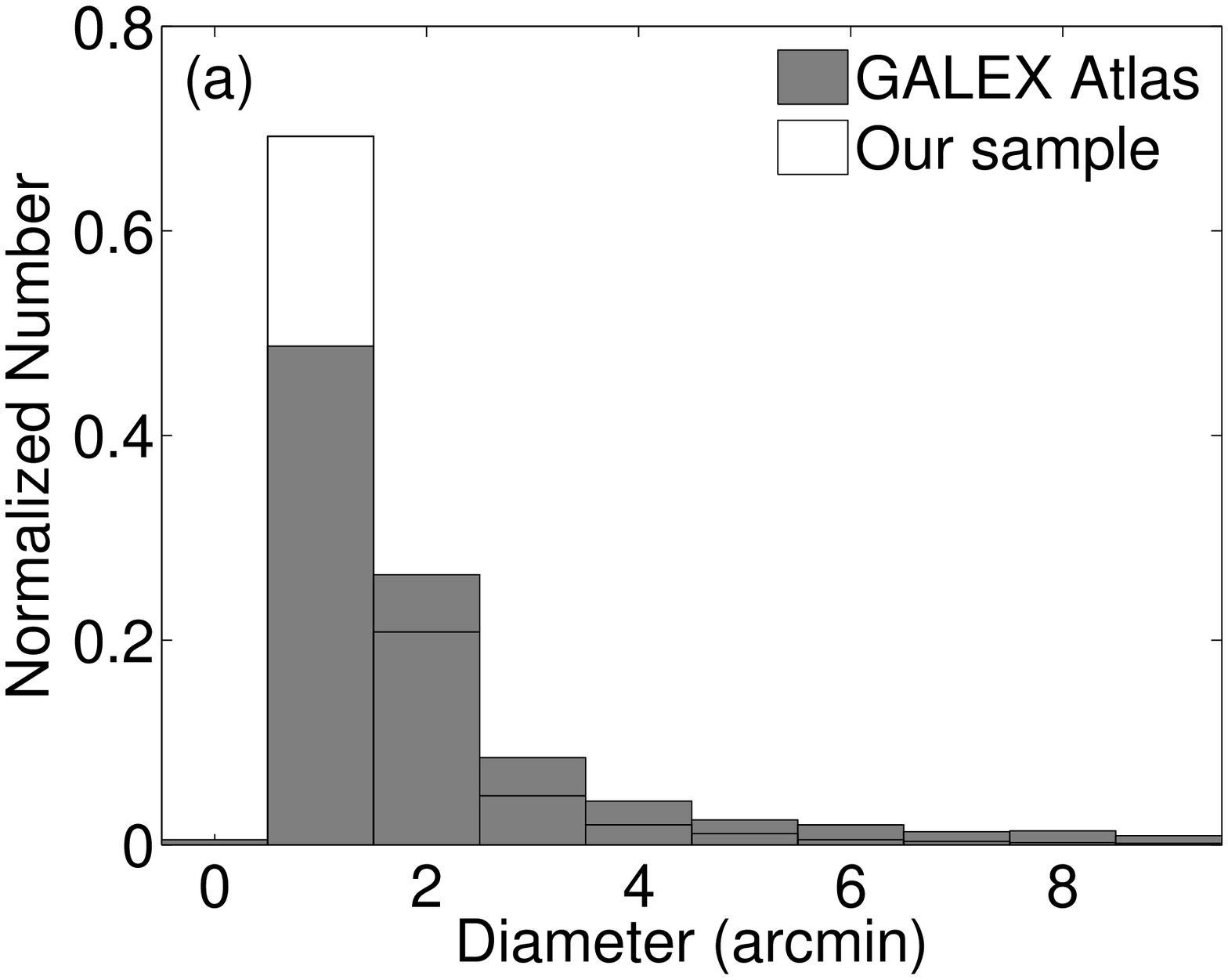}
\includegraphics[width=45mm]{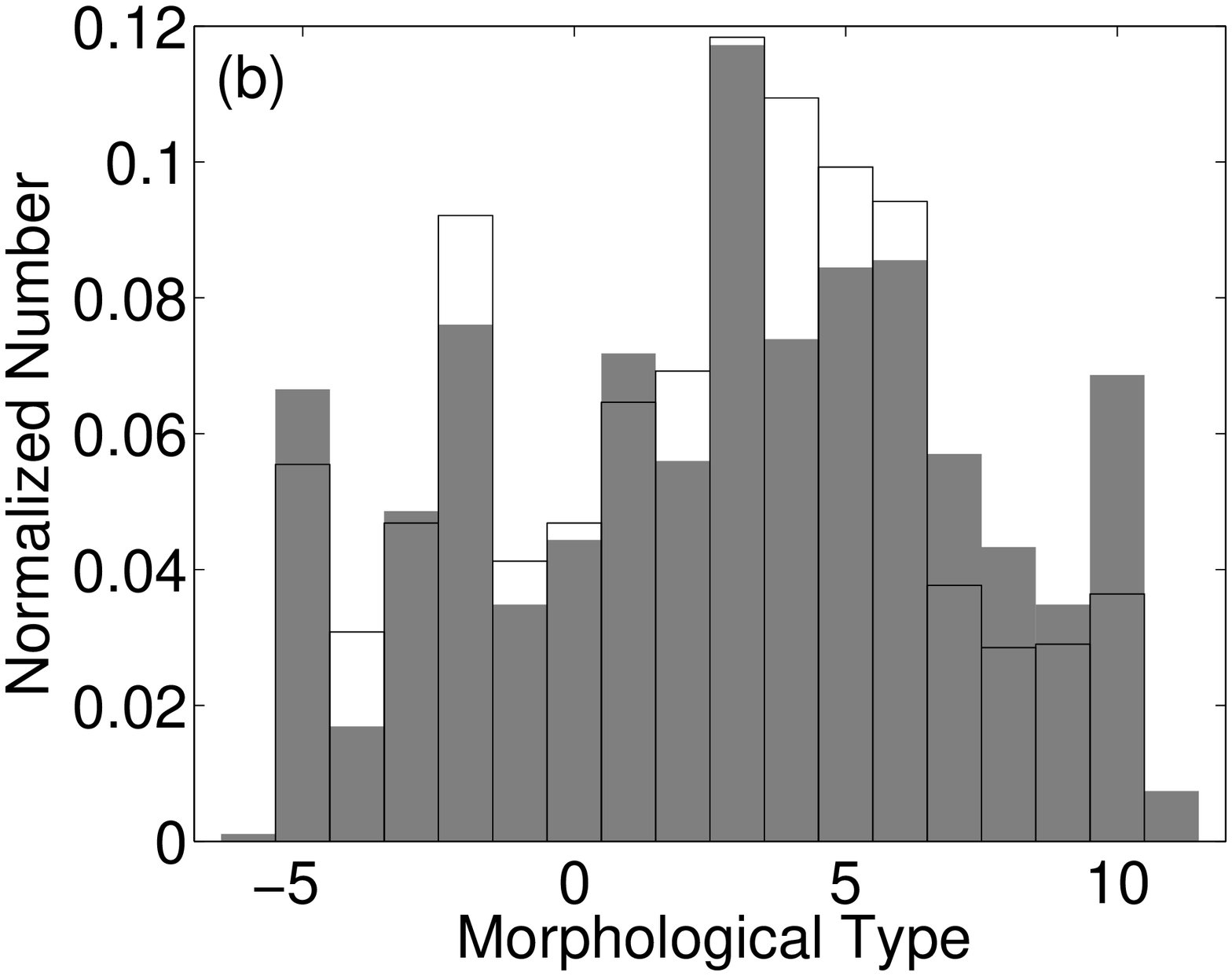}
\includegraphics[width=47mm]{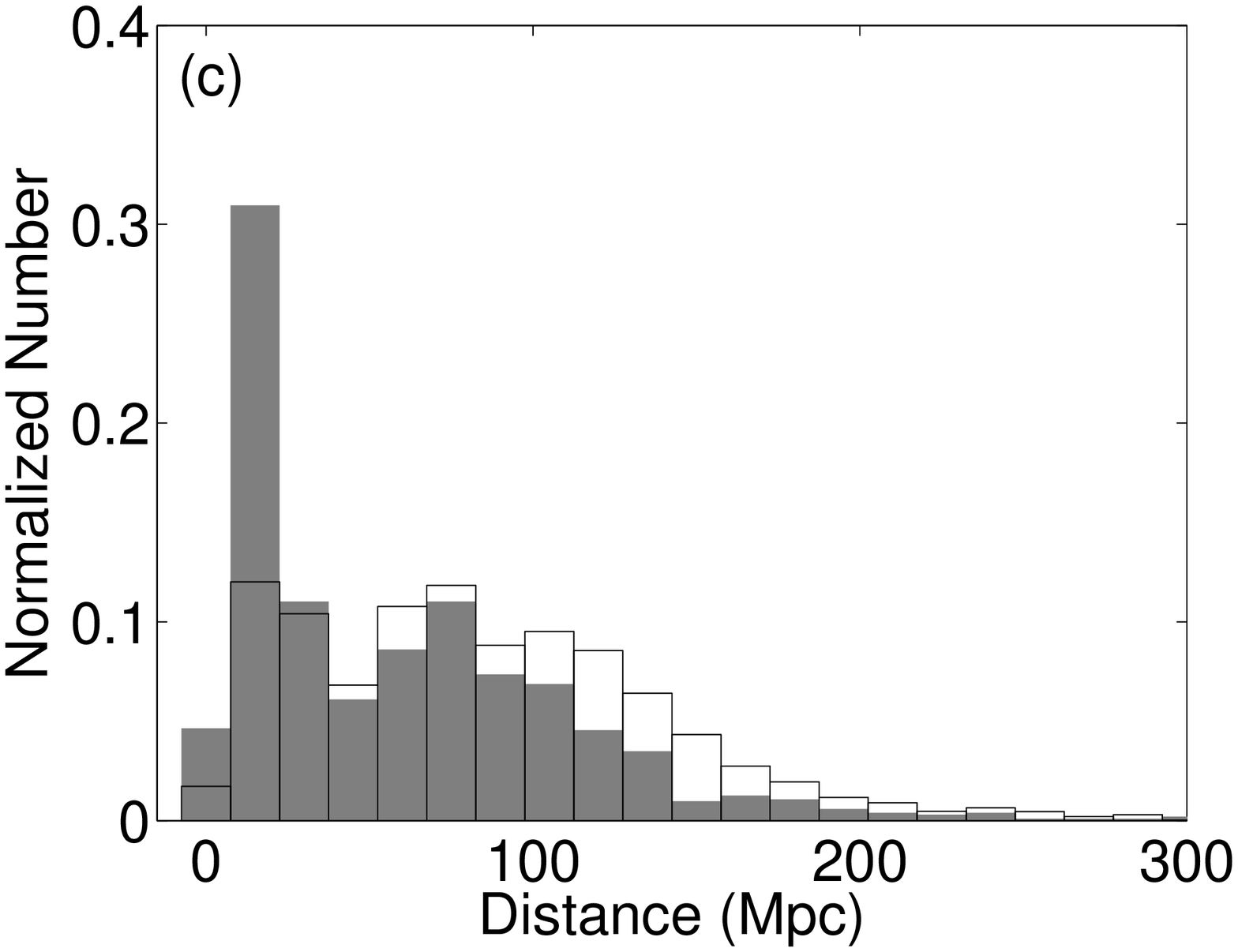}
\caption{Comparison between the \textit{GALEX} Atlas sample (grey-shade histogram) and
our sample (solid line). (a) Major-axis diameters in arcmin. (b) Morphology type. (c) Distance in Mpc.
\label{fig2}
}
\end{figure}

Figure \ref{fig2} shows the normalized distributions of D25 diameter, morphological type, and
distance in \textit{GALEX} Atlas and our sample. By contrast, our sample gives more galaxies with
smaller diameters, earlier types, and larger distances.

\section{Surface Brightness Measurement} \label{sbm}
\textit{GALEX} provides sky background subtracted images, but the background
near the galactic center is overestimated, especially for galaxies with large apparent sizes.
We introduce a new procedure to better estimate the sky background, which is quite
similar to the method in \citet{Zou11}. SExtractor \citep{Bertin96} is used to separate
source pixels from background pixels. The source pixels, together with the region with the galactocentric
distance $r < 2 \times $D25, are masked. Then, the remaining pixels are fitted with
a two dimensional polynomial function to generate the background map. Finally, the fitted
map is subtracted from the intensity image.

Foreground stars contaminate the galaxy flux and need to be masked in the intensity map.
We get the point sources (mostly stars) within the image from \textit{GALEX} photometric
catalogs. The mask aperture is determined by the growth curve, where the corresponding
aperture magnitude is close to the total magnitude presented in the catalog. However, the
catalogued point sources might include some star formation regions, whose UV color is much bluer
than normal stars. Since the number of star formation regions is strongly dependent on the galaxy
type, we separate our entire sample into two subsamples according to the morphological
type: one with T $\geq$ $-$0.5 identified as spiral/irregular galaxies (types from $Sa$ to $Irr$)
and the other with T $<$ $-$0.5 identified as elliptical/lenticular galaxies (types from $E$
to $S0a$).
For early-type galaxies, all point sources detected by \textit{GALEX} are regarded
as foreground stars and masked.
For late-type galaxies, we only mask the point sources with (FUV $-$ NUV) $>$ 1,
and keep bluer point sources with (FUV $-$ NUV) $<$ 1 as star forming regions.
The nuclei of galaxies, also point-like, are not masked.
For galaxies with single UV-band observations, we define foreground stars by visually
inspecting both the UV image and the color image of the Digitized Sky Survey.

\begin{figure}
  \centering
\includegraphics[width=35mm]{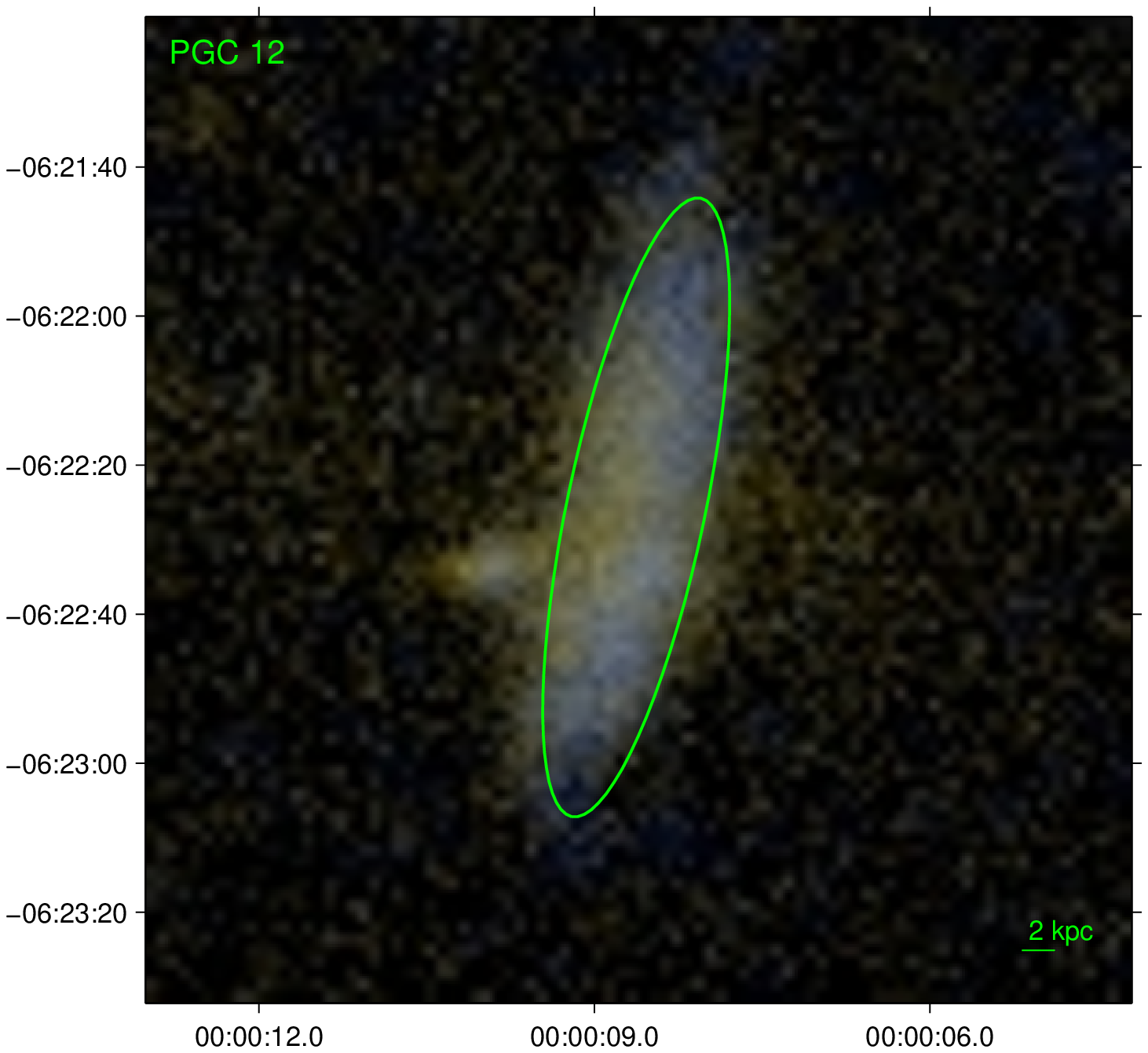}
\includegraphics[width=40mm]{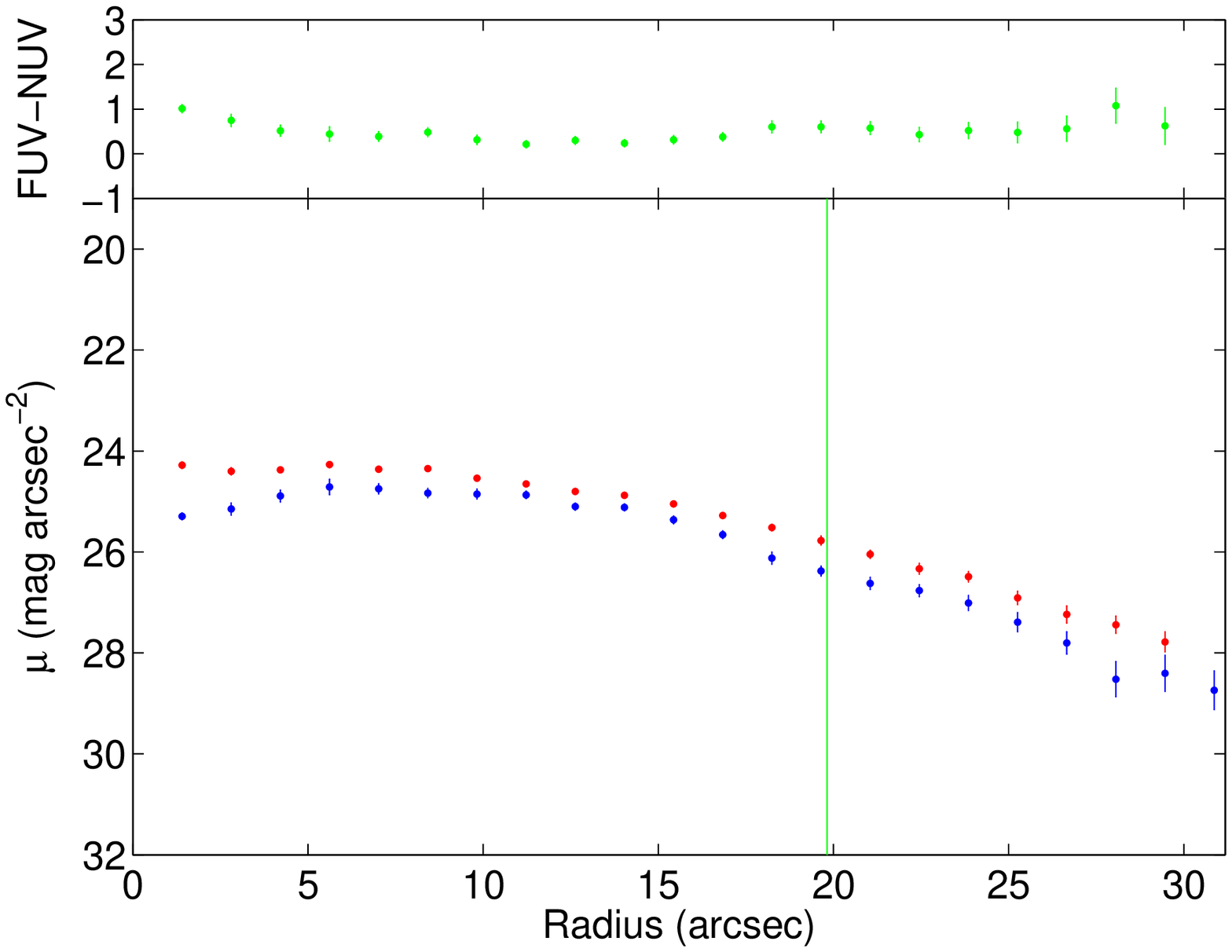}
\includegraphics[width=35mm]{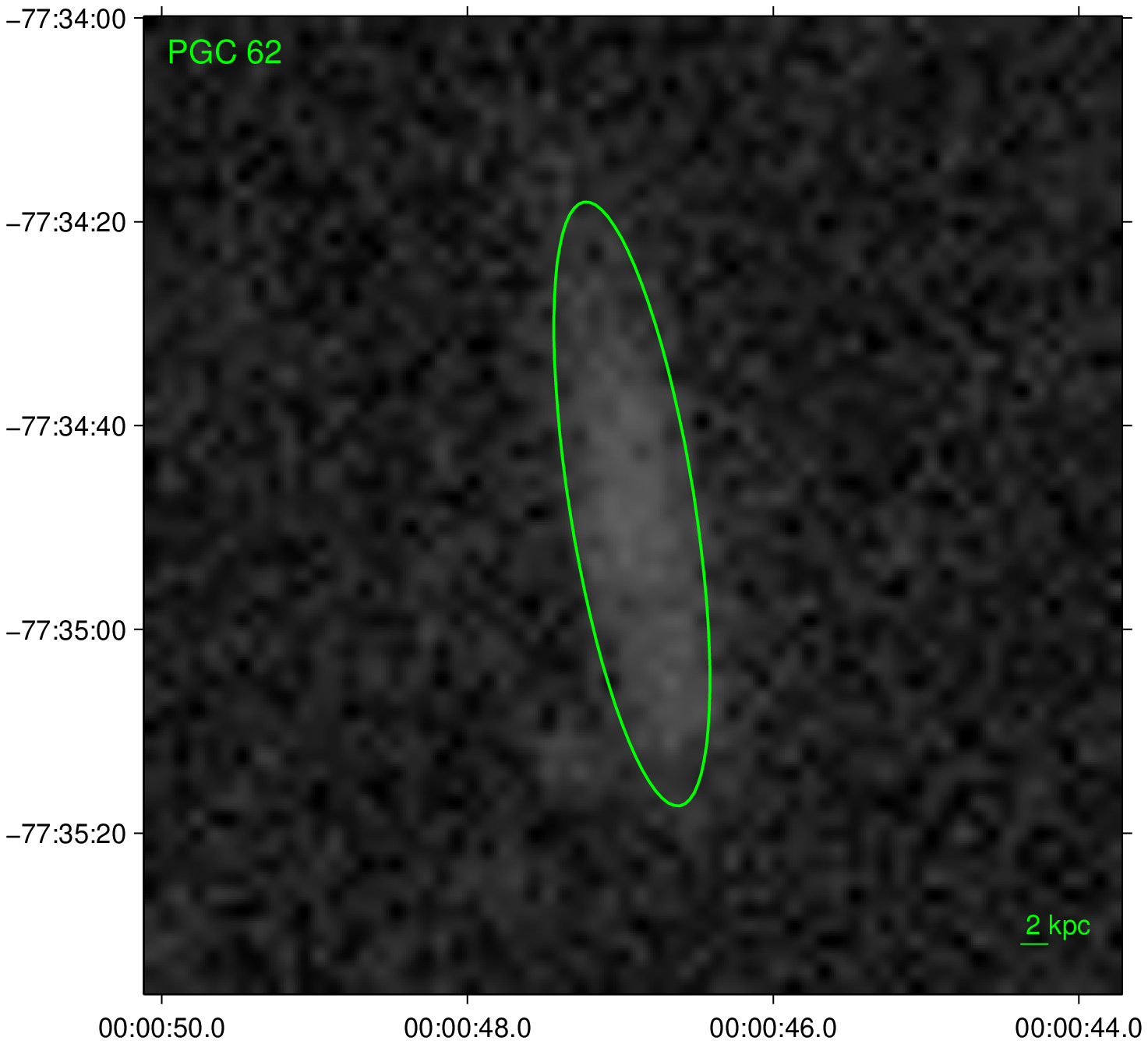}
\includegraphics[width=40mm]{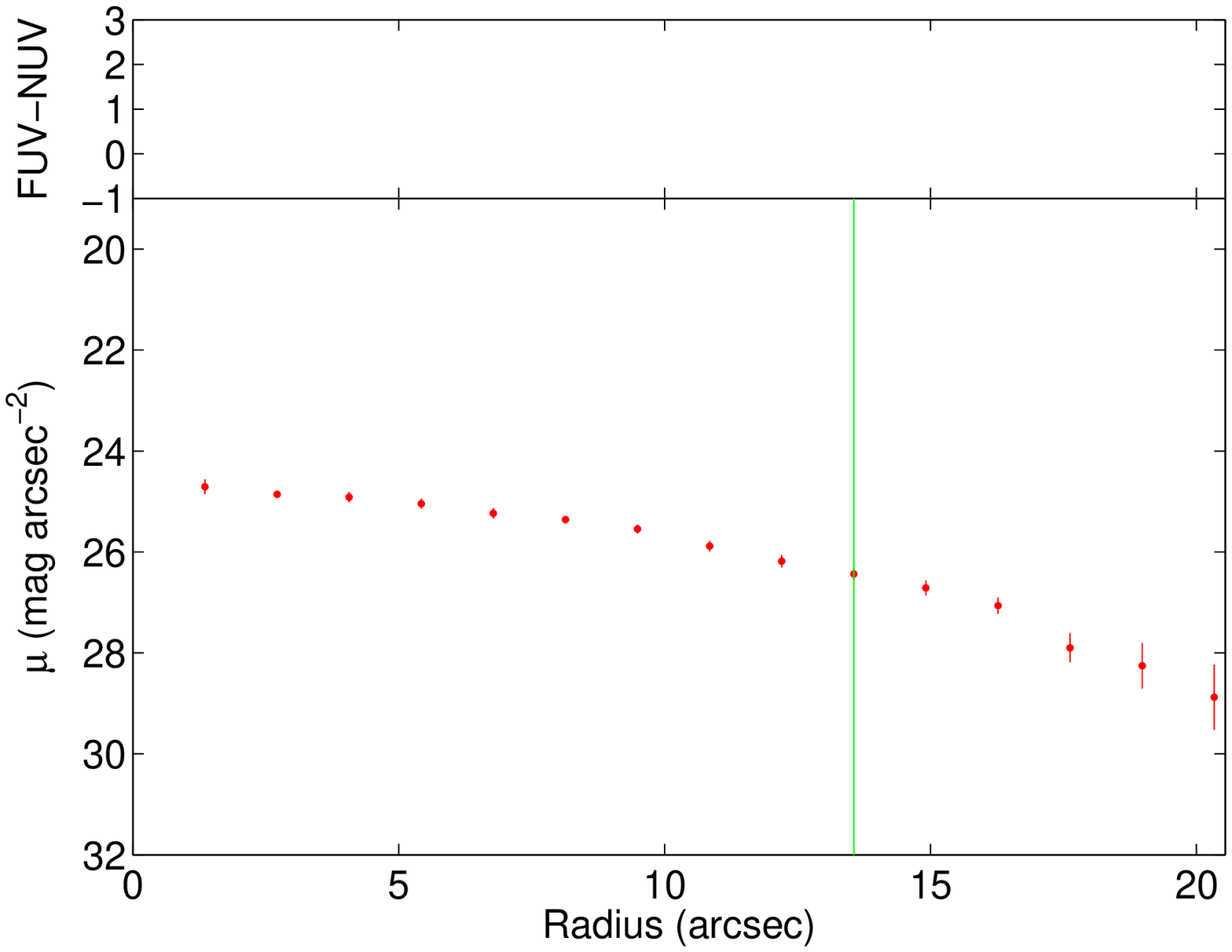}
\includegraphics[width=35mm]{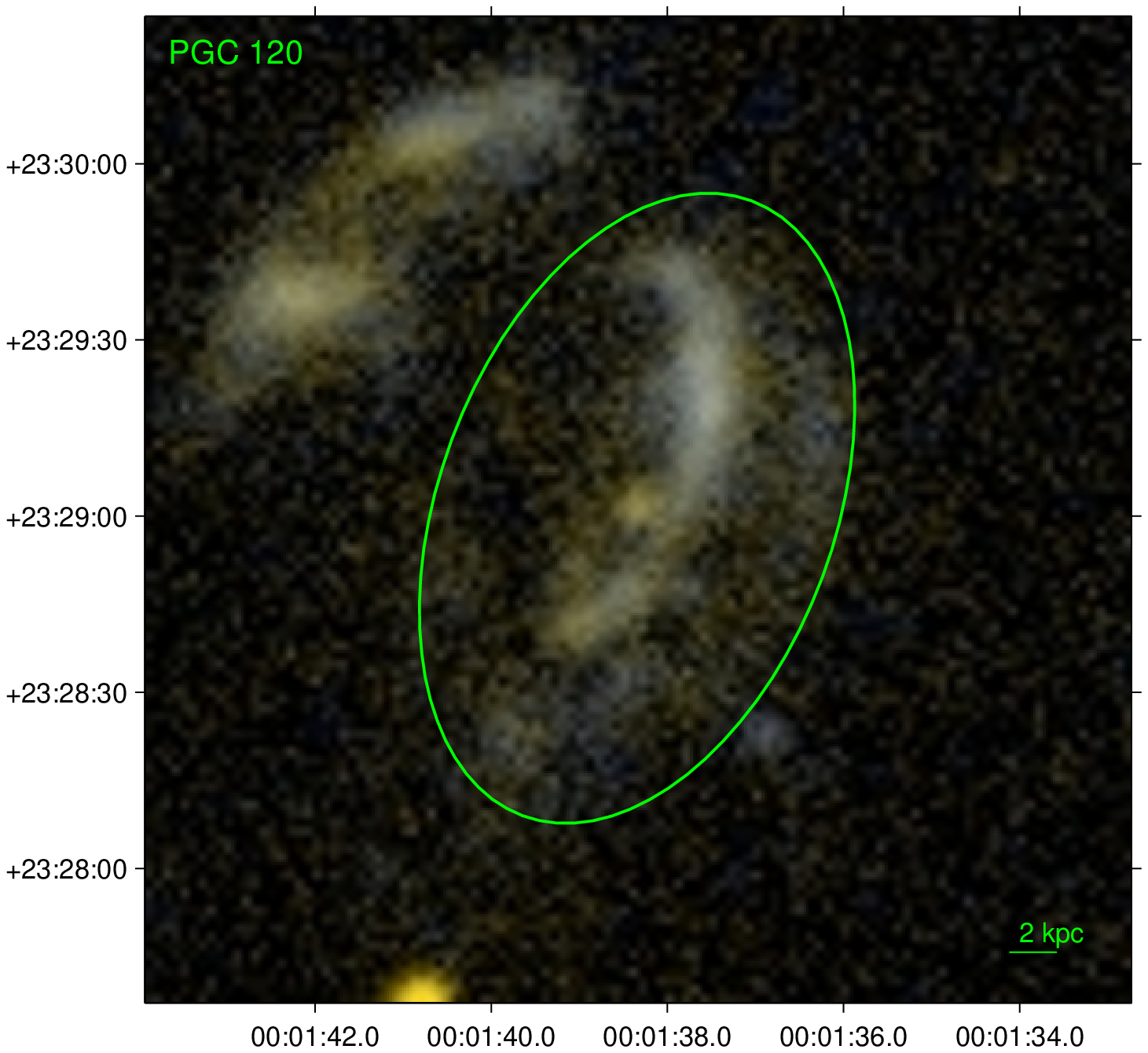}
\includegraphics[width=40mm]{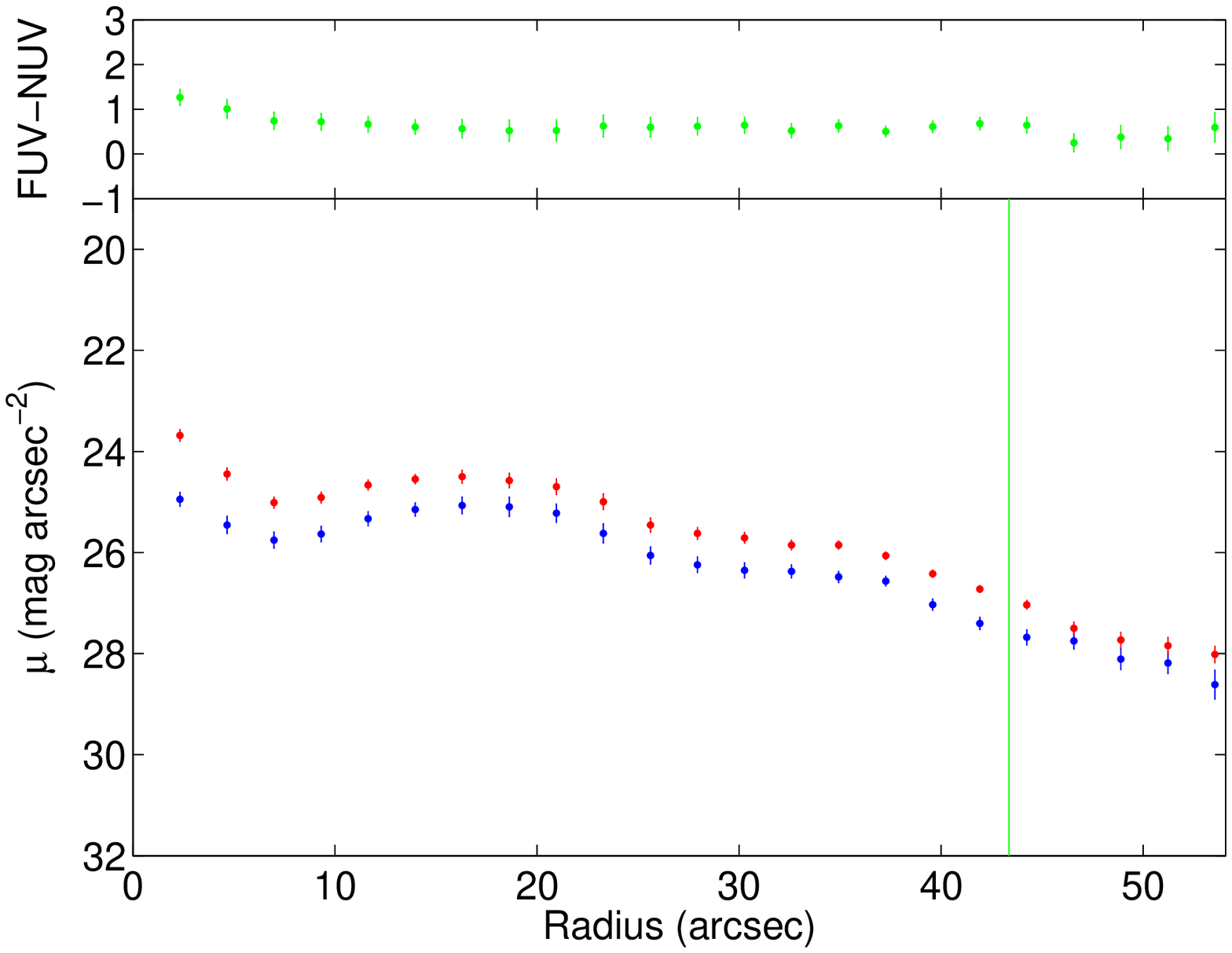}
\includegraphics[width=35mm]{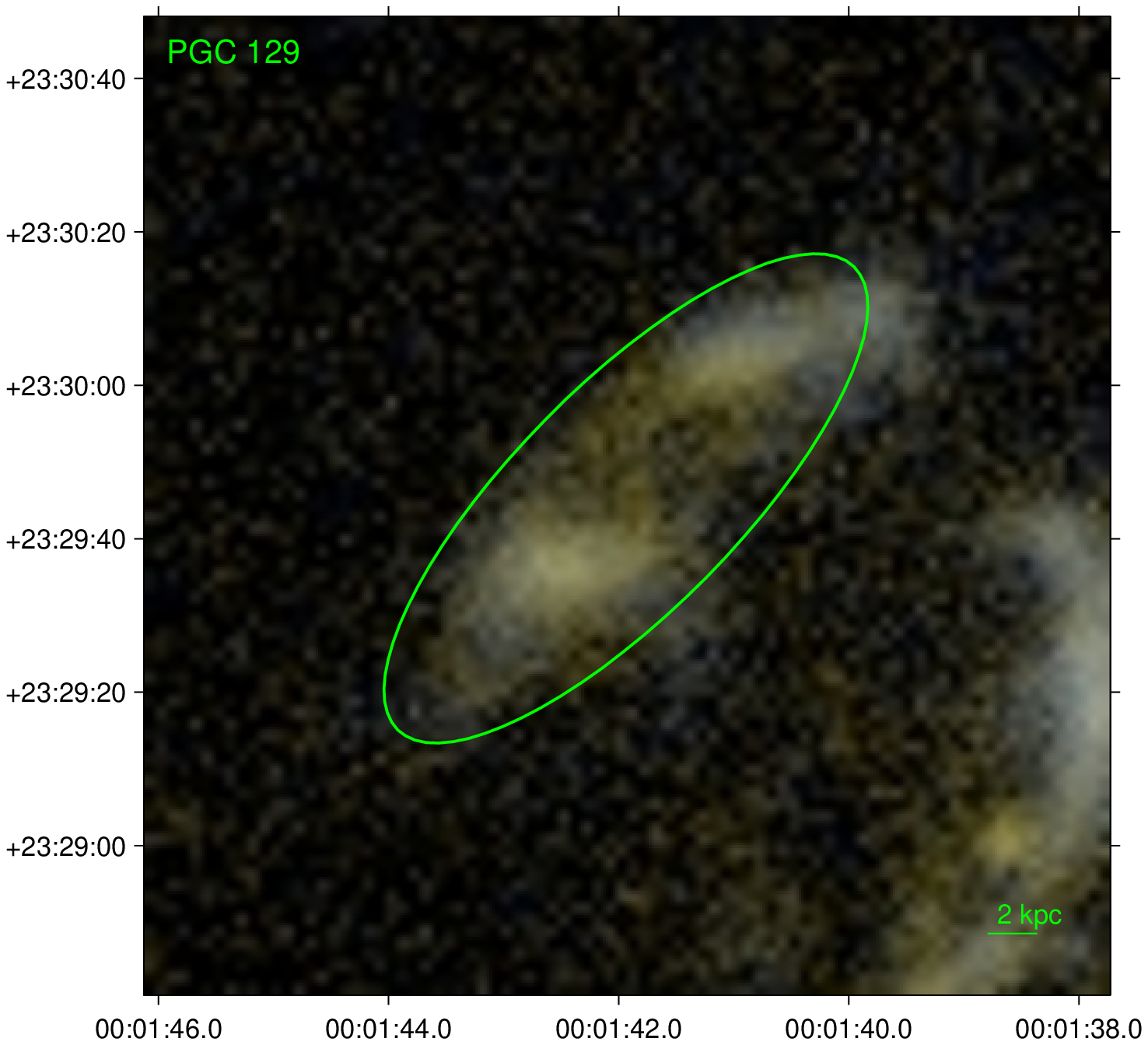}
\includegraphics[width=40mm]{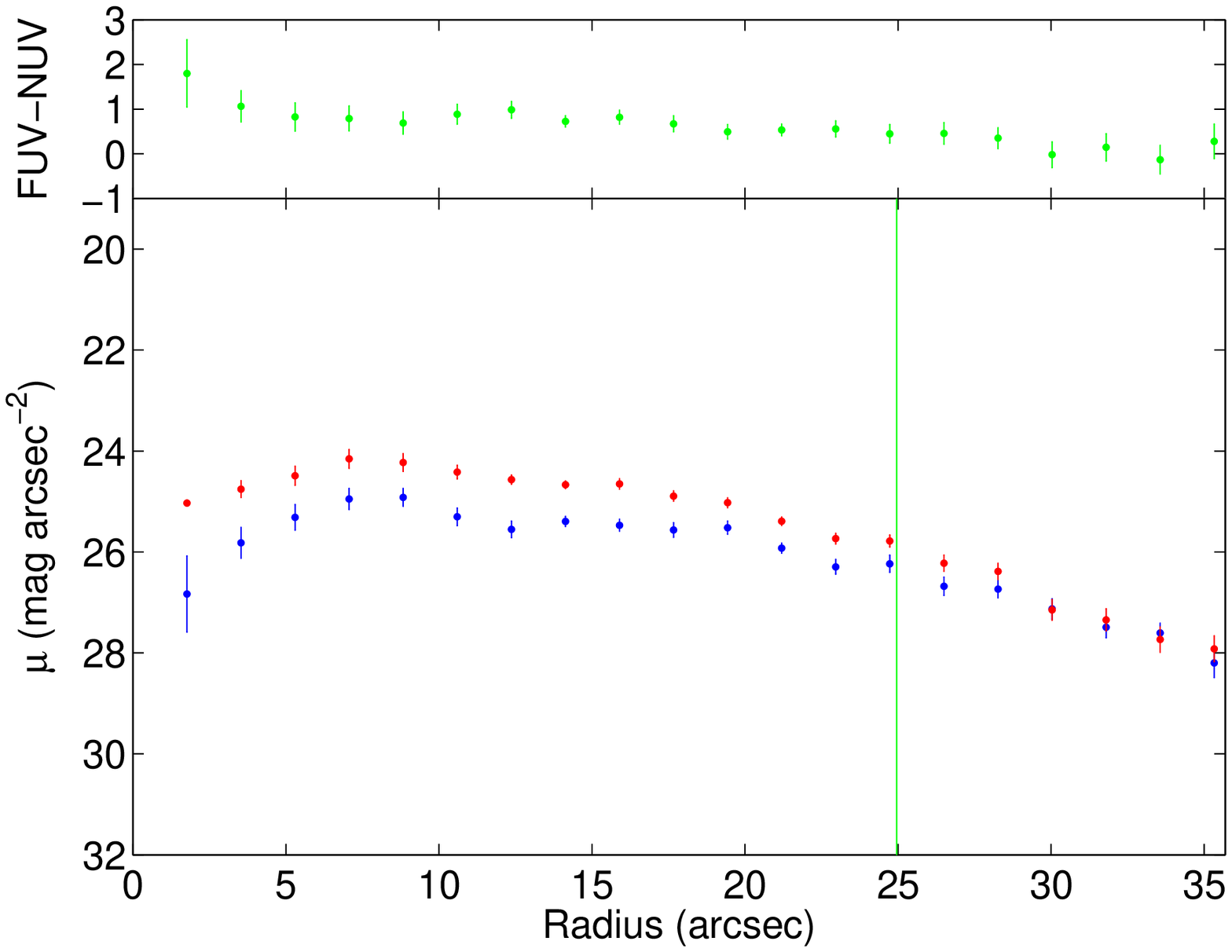}
\includegraphics[width=35mm]{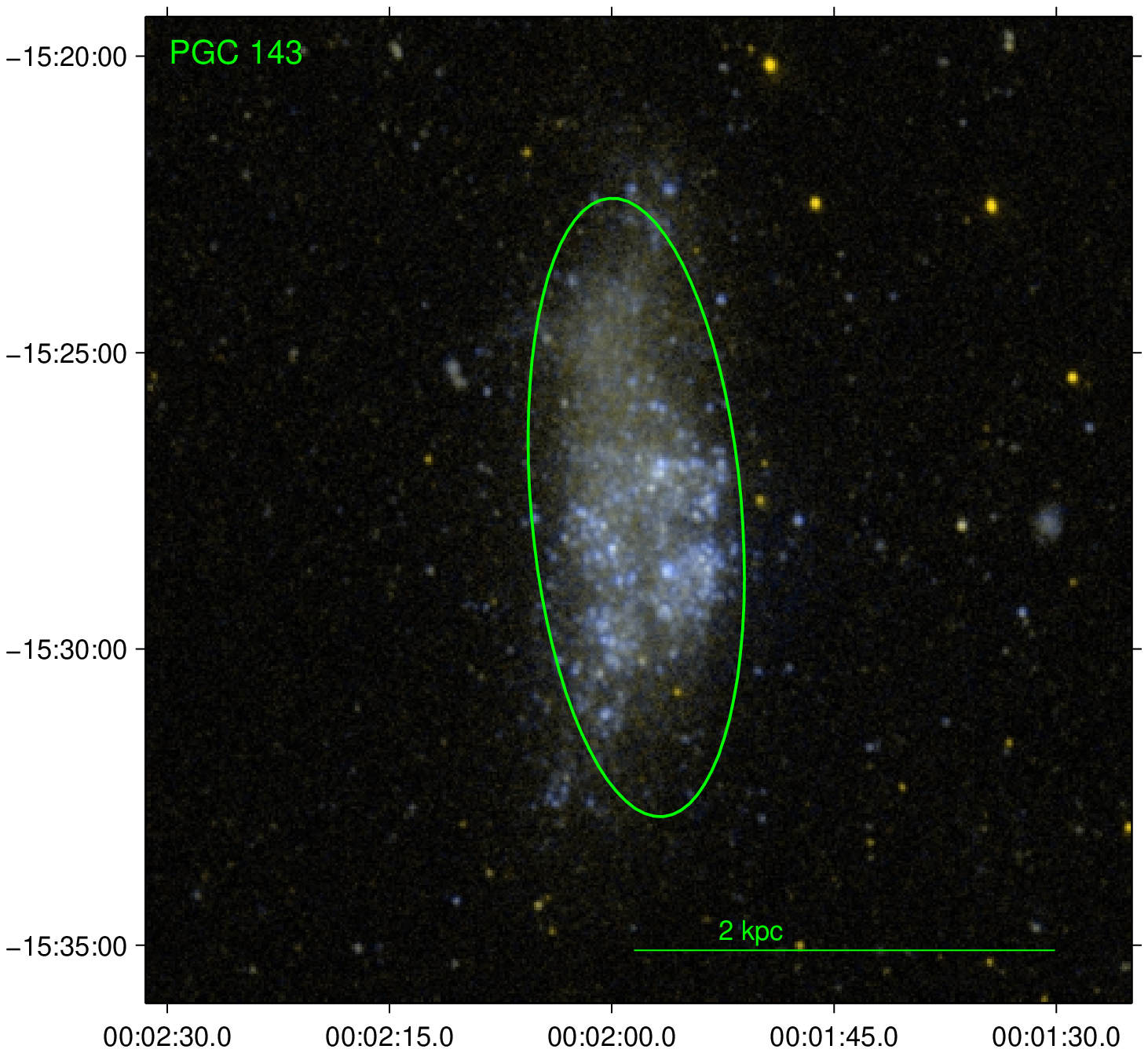}
\includegraphics[width=40mm]{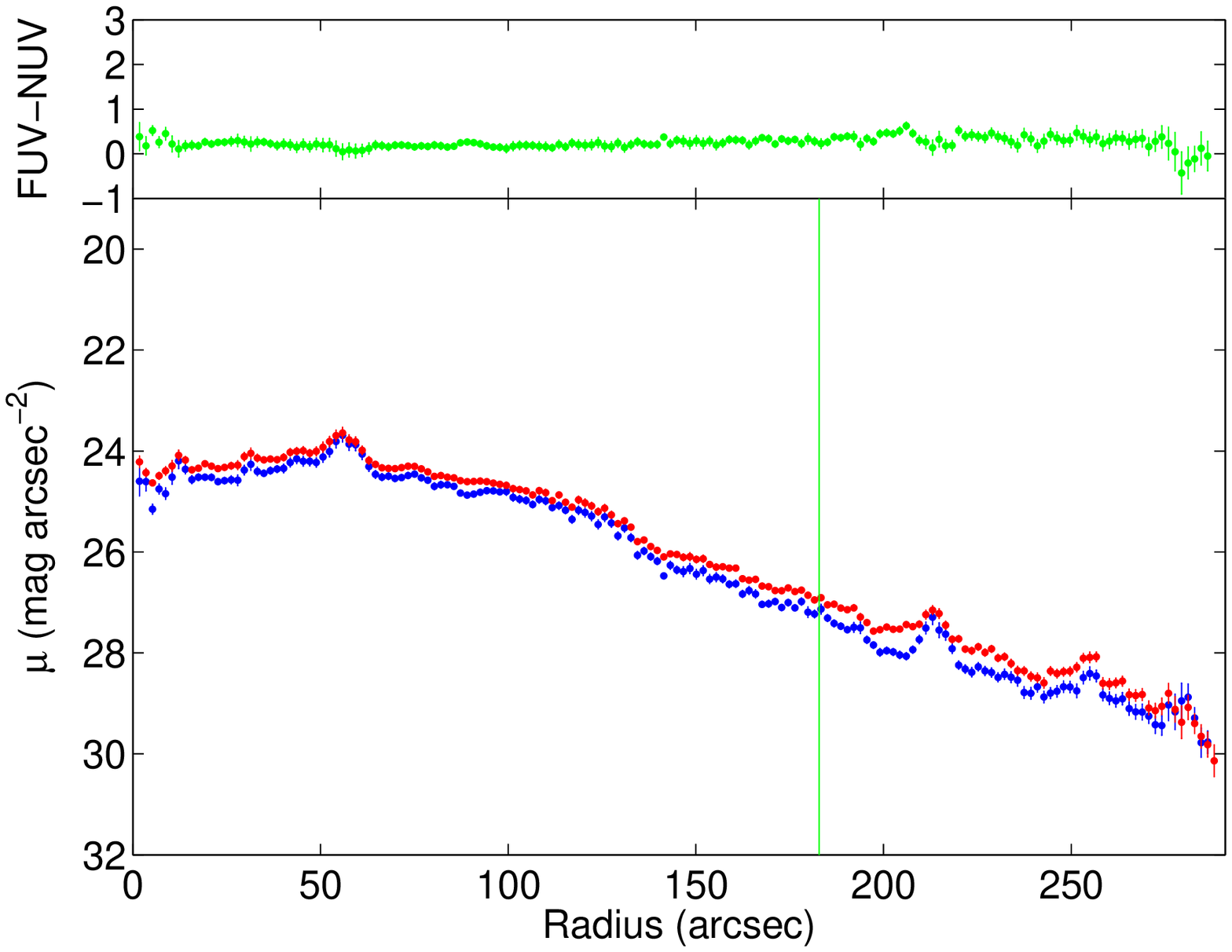}
\includegraphics[width=35mm]{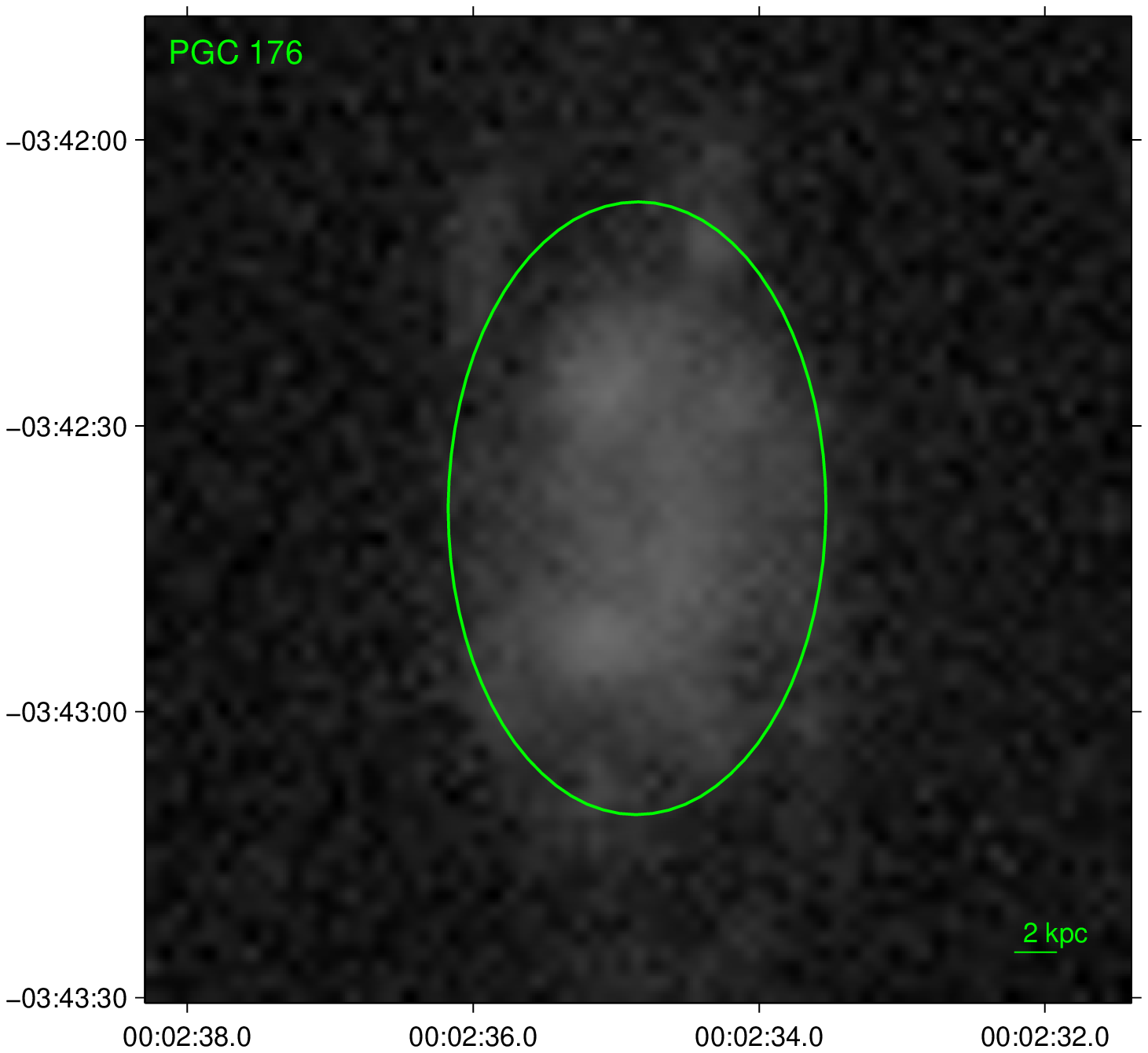}
\includegraphics[width=40mm]{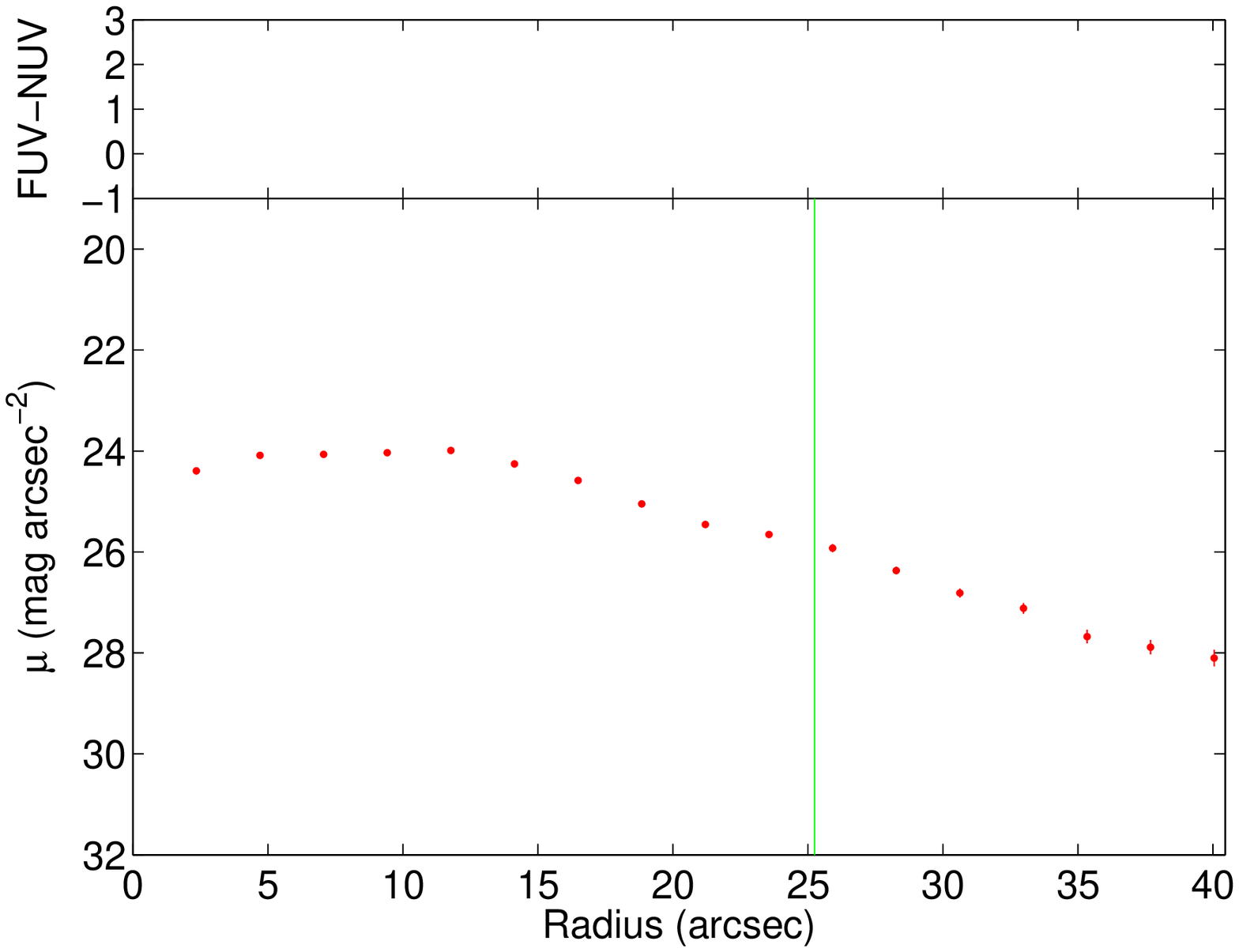}
\includegraphics[width=35mm]{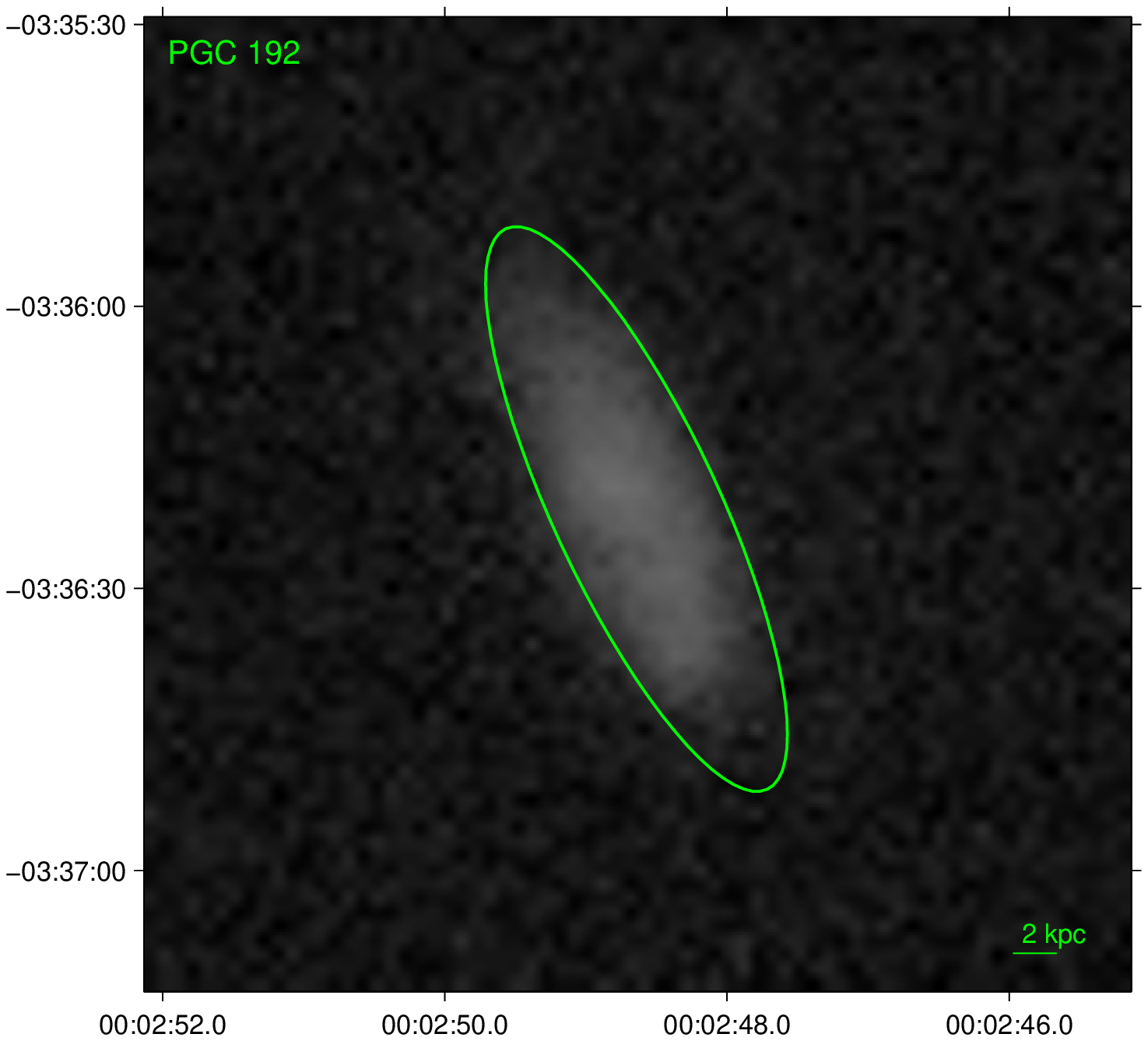}
\includegraphics[width=40mm]{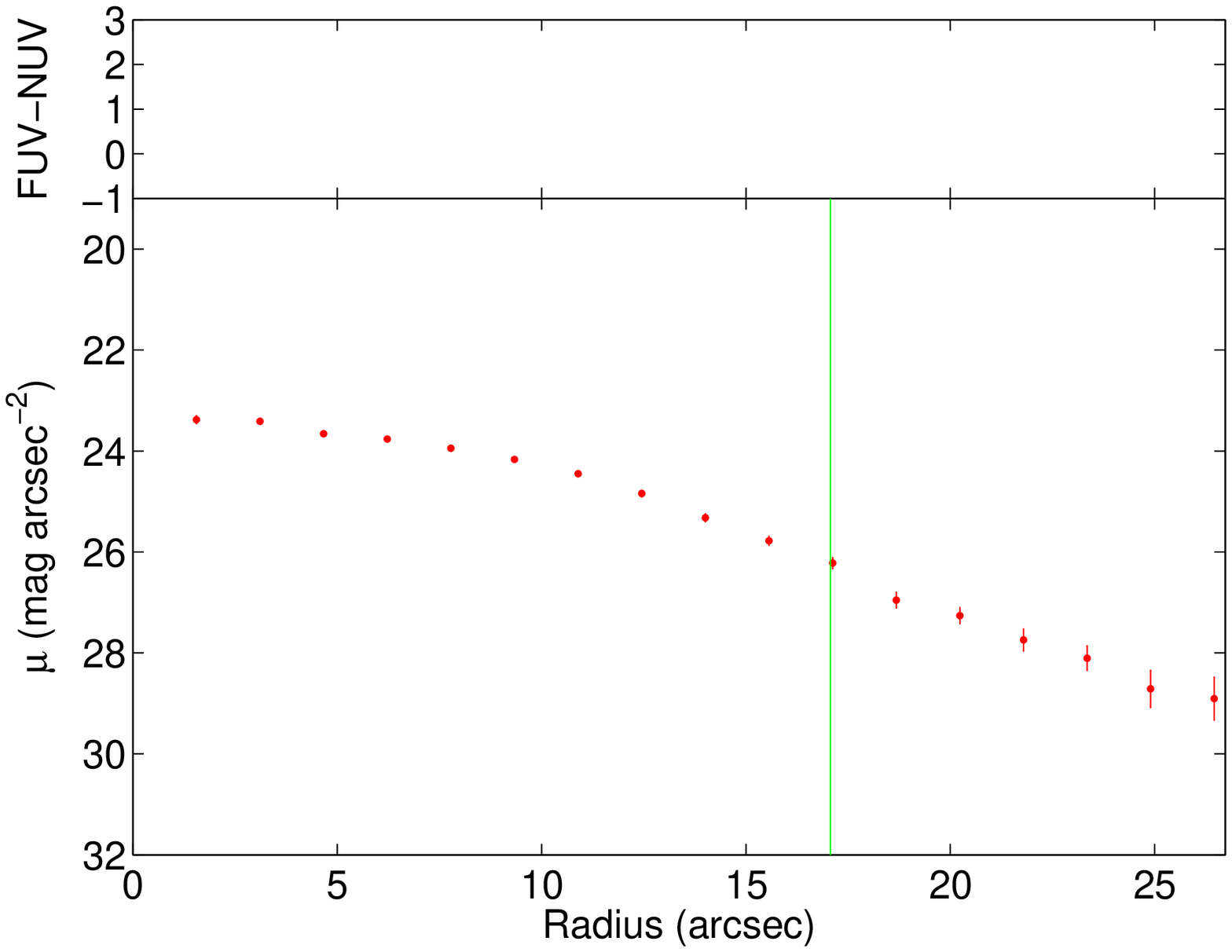}
\includegraphics[width=35mm]{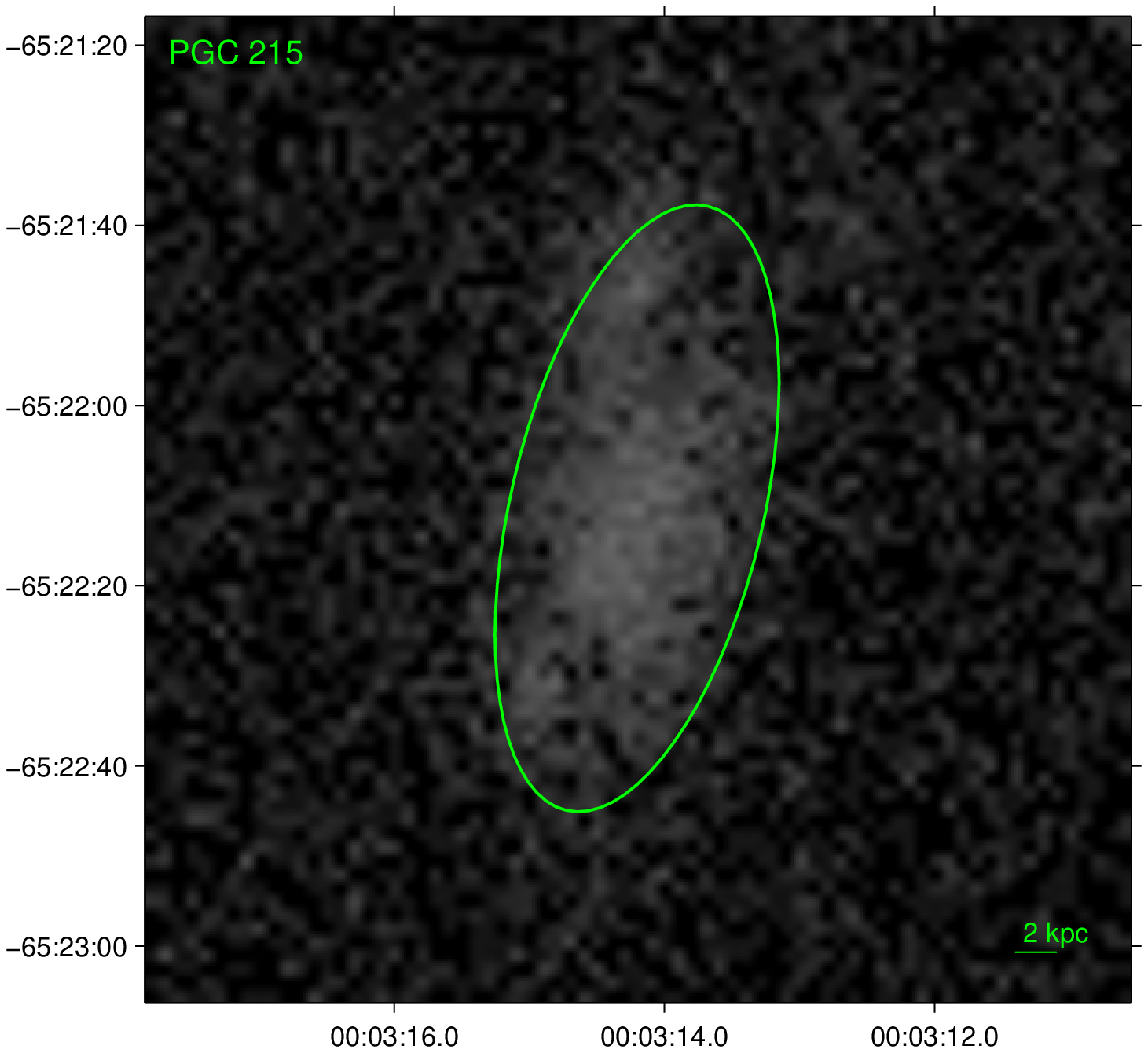}
\includegraphics[width=40mm]{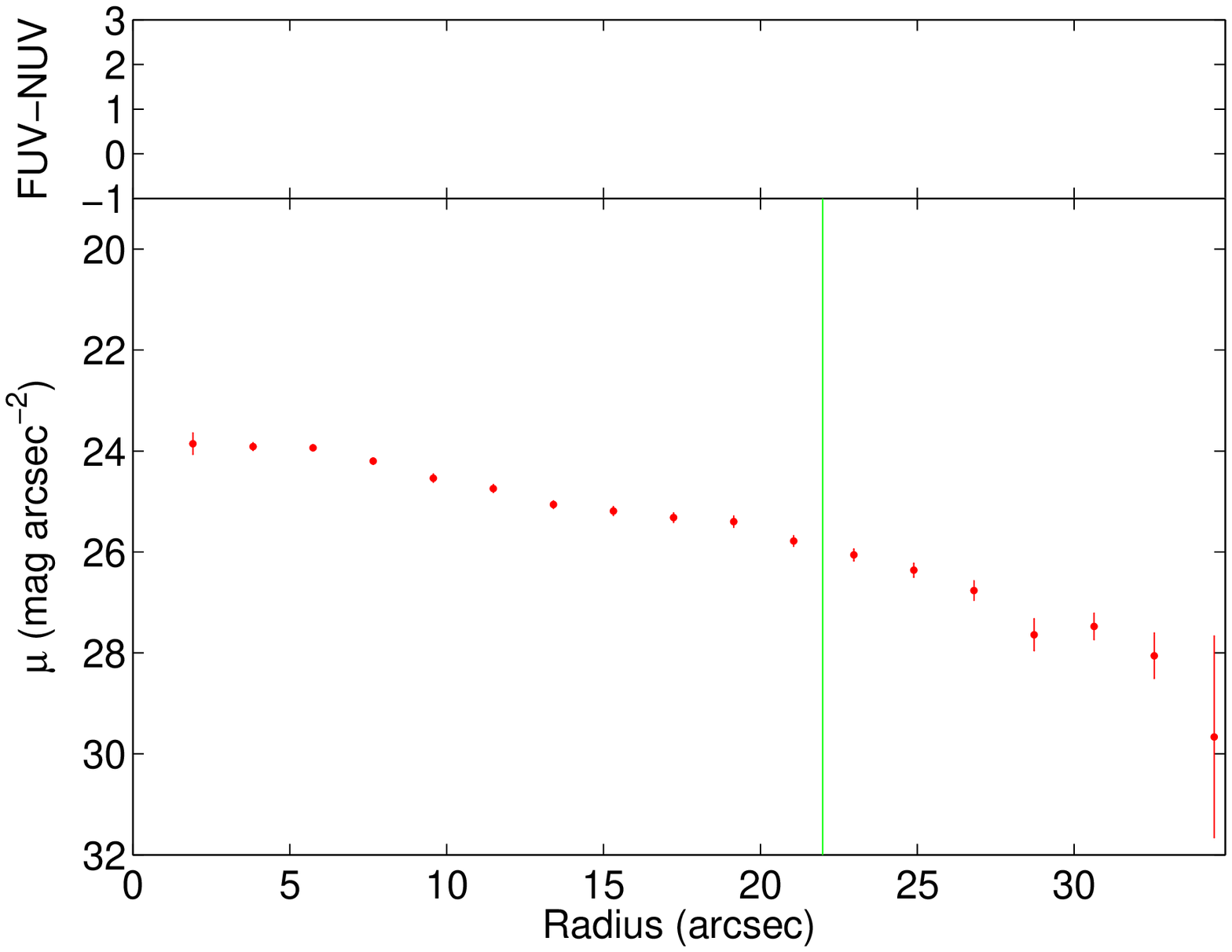}
\includegraphics[width=35mm]{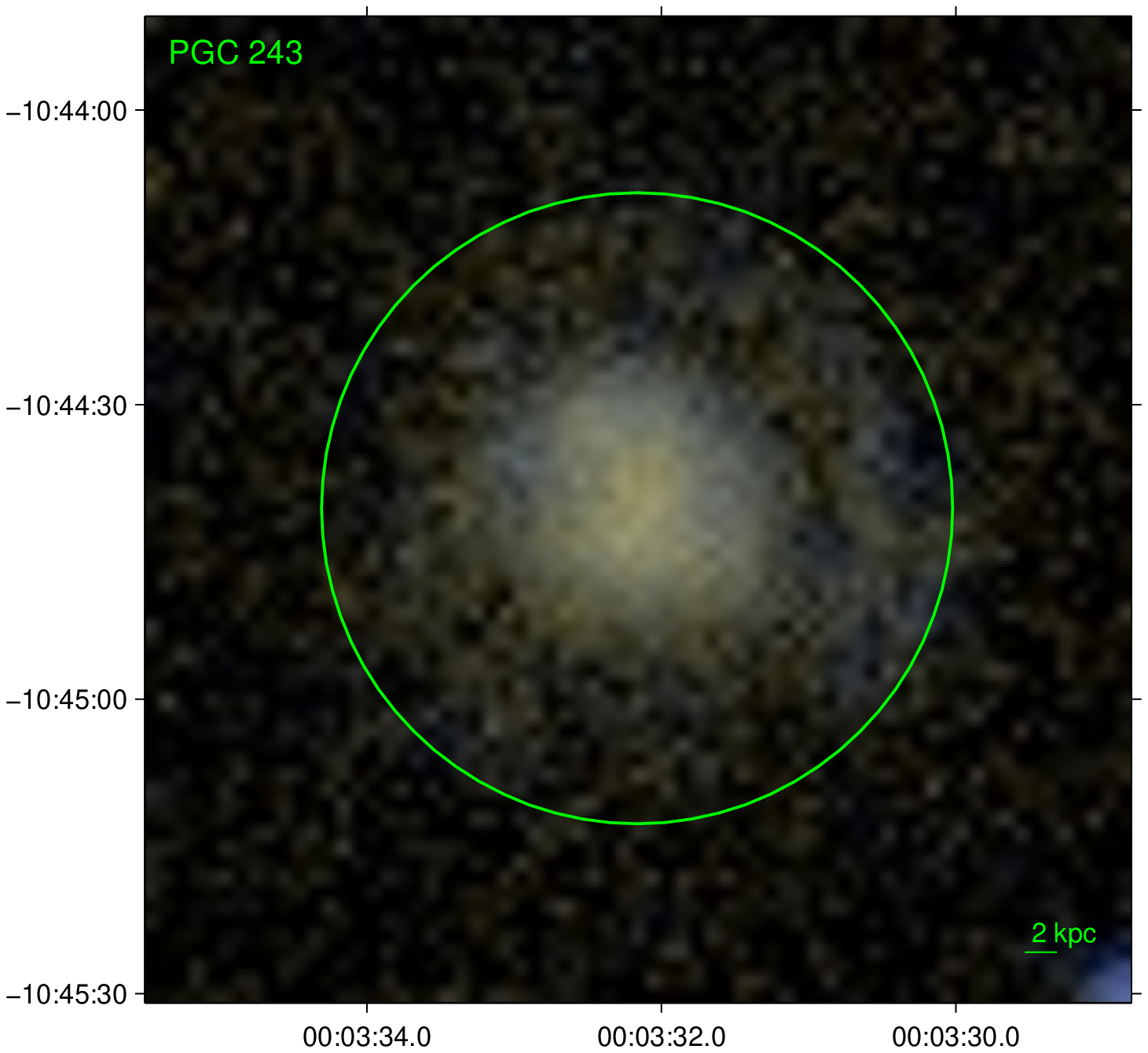}
\includegraphics[width=40mm]{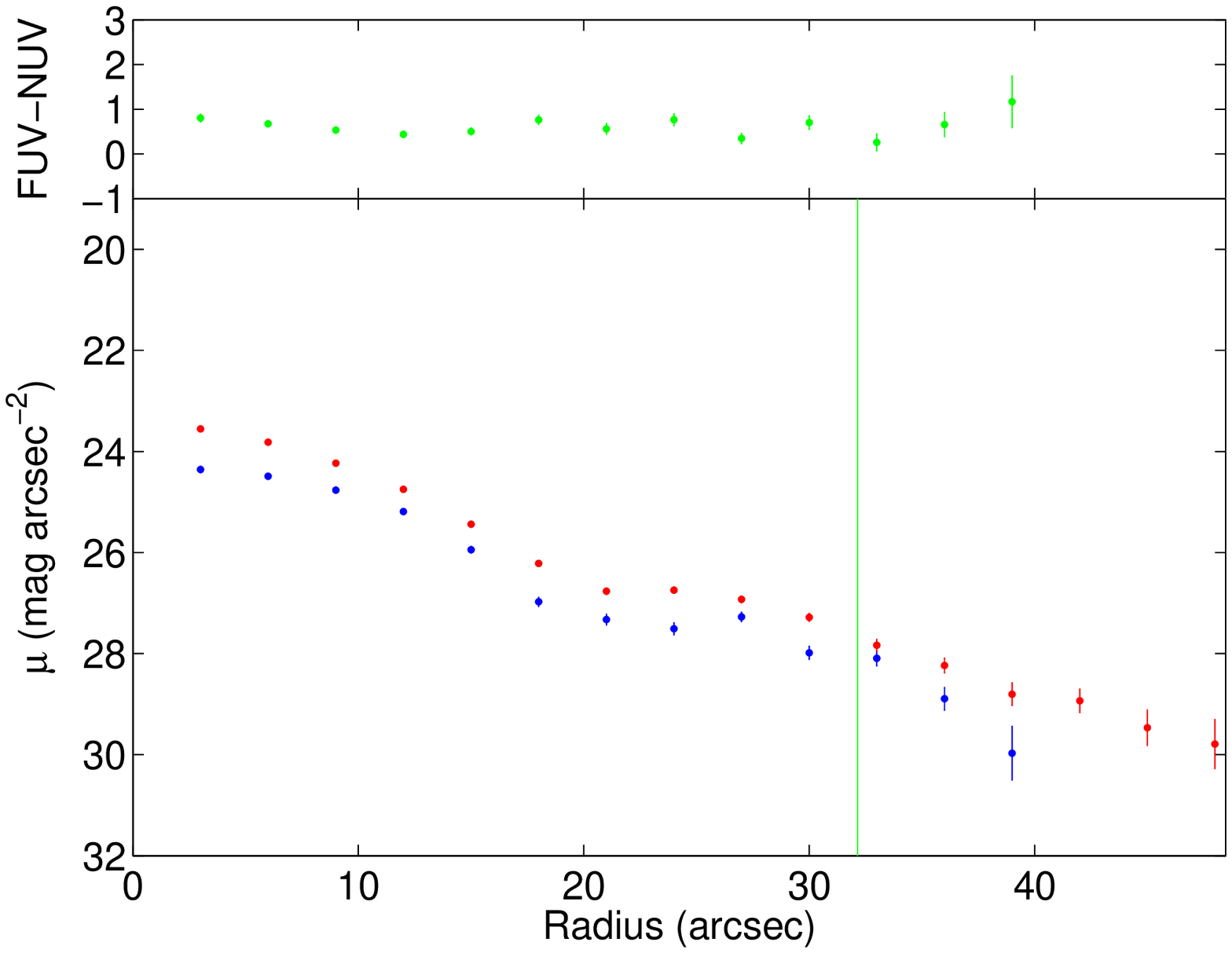}
\includegraphics[width=35mm]{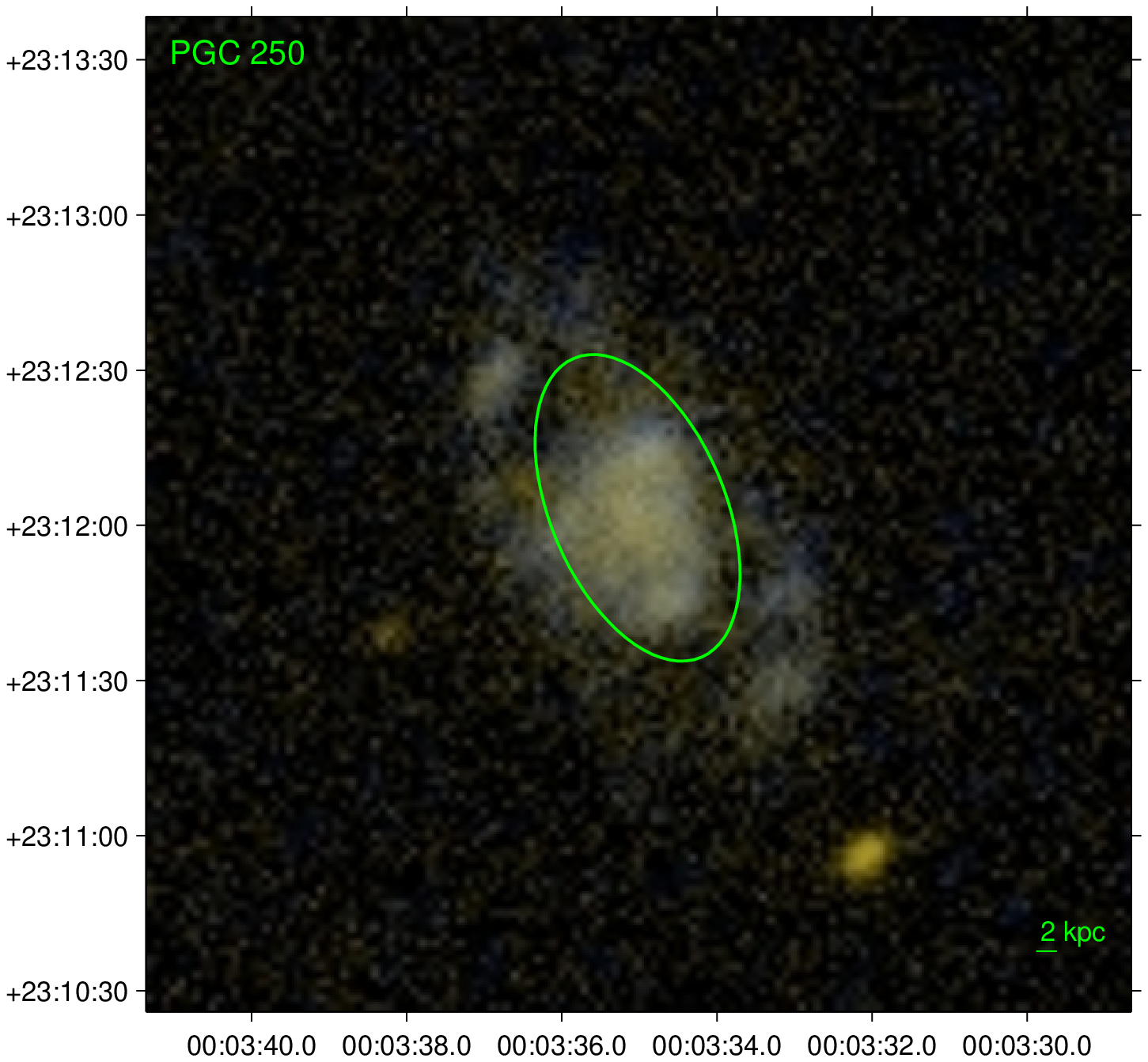}
\includegraphics[width=40mm]{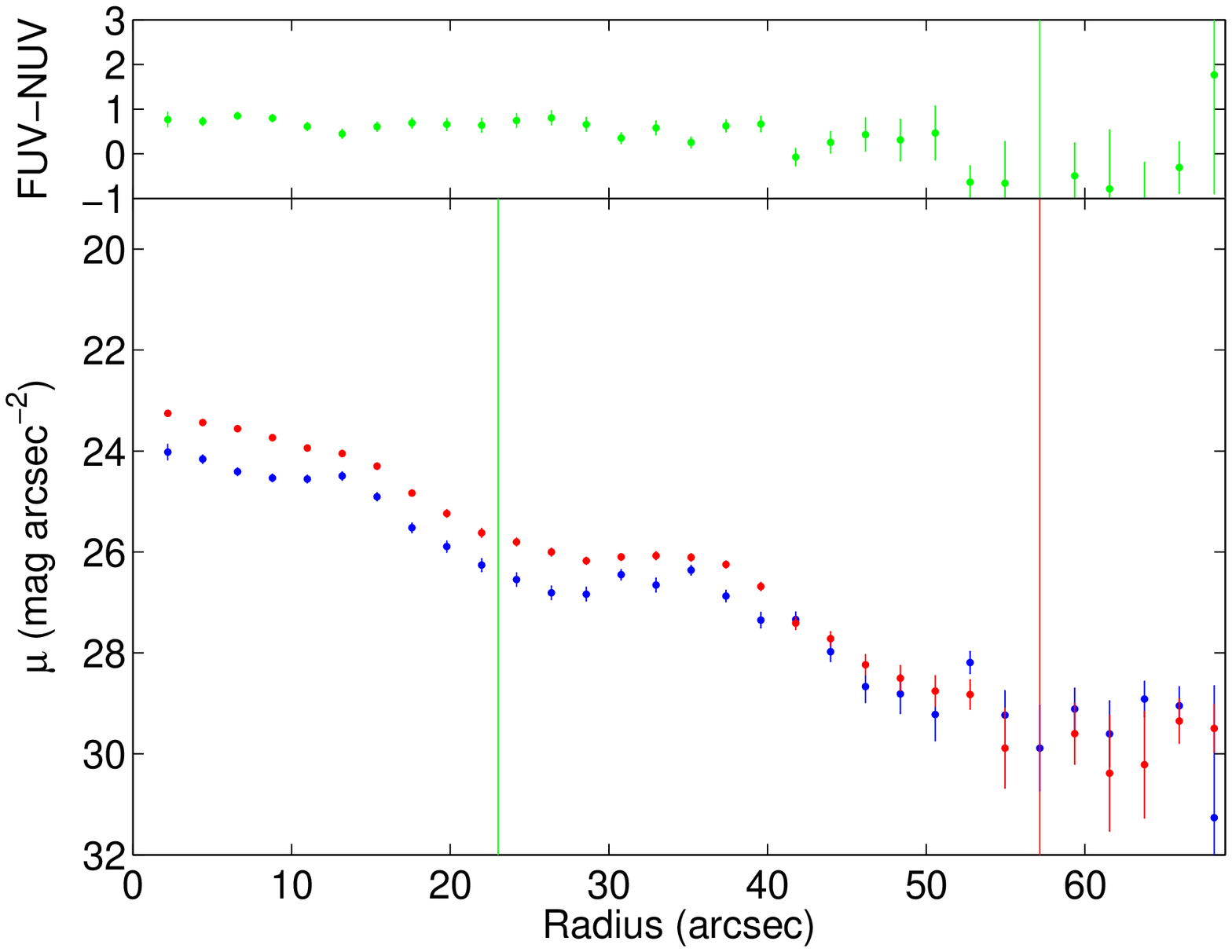}
  \caption{Left: False color images of our \textit{GALEX} nearby galaxies. R.A. and Dec. are in J2000.0. The D25 ellipse is drawn in green. Right: radial surface brightness and color profiles. Red, blue, and green points present the NUV, FUV and (NUV $-$ FUV) profiles, respectively. The green vertical line is the equivalent radius corresponding to D25. \label{fig3}}
\end{figure}

We then use the MATLAB package of astronomy and astrophysics for the measurement \citep{Ofek14}. 
The radial surface brightness profile is computed with a series of elliptical annuli, whose
major radii range from 3$\arcsec$ to at least 1.5 times the D25 radius. The brightness
errors are estimated with the method in \citet{Gil05}. Figure \ref{fig3}
shows some examples of the false color images and surface brightness profiles of galaxies
in our sample. The radial (FUV $-$ NUV) color profile is also plotted.

\section{The Catalog Content} \label{cc}
\subsection{Asymptotic and D25 Magnitudes}
We adopt the technique introduced by \citet{Cairos01} to calculate the asymptotic
magnitudes. First, the accumulated flux and the gradient of the growth curve
at each radius are computed. Then, an appropriate radial range away from the
galaxy center is visually chosen, where an error-weighted linear fit to the relation of
accumulated flux vs. the gradient of the growth curve is performed.
Finally, the intercept of this linear
fit is regarded as the asymptotic magnitude, and corresponding magnitude error
is derived from the error of the linear fit.

We also calculate the aperture magnitudes inside the D25 ellipses. The errors
are estimated from the random noise inside the apertures and the uncertainties
of the sky background subtraction (see \citealt{Gil05}). All the errors do not
include the uncertainty of NUV and FUV photometric zeropoints, which are about 0.15 and
0.09 mag, respectively.

\subsection{Effective surface brightness and concentration indices}
Based on the growth curve, we derive the radii containing 50\%, 20\%, 25\%,
75\%, 80\% of the total luminosity ($r_{\mathrm{eff}}$,$r_{20}$, $r_{25}$,
$r_{75}$, and $r_{80}$, resepectively). The effective surface brightness is
calculated as $I = 0.5L_{\mathrm{FUV}}/\pi r^{2}_\textrm{eff}$ and the
concentration indices are calucated as C31=$r_{75}$/$r_{25}$ and
C42 = 5log($r_{80}$/$r_{20}$) \citep{de77,Kent85}.

\subsection{Archival optical and infrared data}
Multi-wavelength data of nearby galaxies could improve our understanding of their
physical nature. 
We add some corollary data from other optical and infrared surveys
into our catalog. HyperLeda provides the total $B$ magnitude ( $B_{T}$) and the total
$I$ magnitude ($I_{T}$). A total of 4064 galaxies have $B_{T}$ magnitudes and 3265
galaxies have $I_{T}$ magnitudes. There are 640 and 813 galaxies in our sample with
$(U-B)_{T}$ and $(B-V)_{T}$ colors, respectively. In near infrared (NIR), we compile
$JHK$-band photometry from the 2MASS Large Galaxy Atlas (LGA) \citep{Jarrett03}. If
galaxies are unavailable in LGA, $JHK$ photometry in 2MASS Extended Source
Catalog (XSC) are used. A total of 3708 galaxies have $JHK$ magnitudes. In mid and far
infrared (MIR and FIR), the photometry in 12$\mu$m, 25$\mu$m, 60$\mu$m and 100$\mu$m
are obtained from Infrared Astronomical Satellite (IRAS). We follow the priority
given by \citet{Gil07} to compile the photometry (\citealt{Rice88,Moshir90} and the
$IRAS$ Point Source Catalog). 
The 60$\mu$m and 100$\mu$m are used to estimate the far infrared emission. There are 1570 galaxies
with infrared fluxes in both 60$\mu$m and 100$\mu$m. The UV properties and archival
optical and IR data are presented in presented in Table \ref{tab2} and Table \ref{tab3}, respectively.

\section{Comparison with \textit{GALEX} Atlas} \label{compga}
\begin{figure}
\centering
\includegraphics[width=45mm]{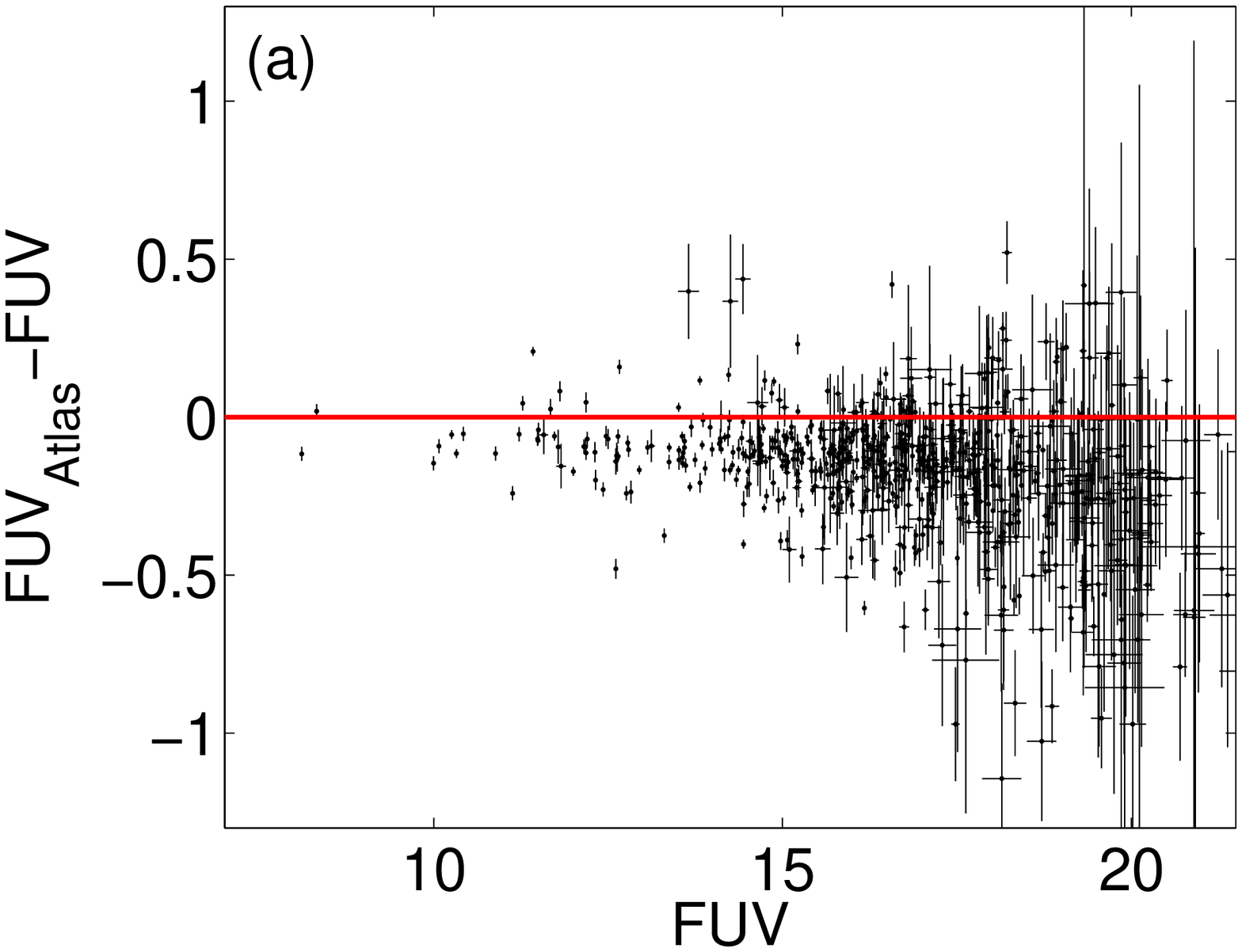}
\includegraphics[width=45mm]{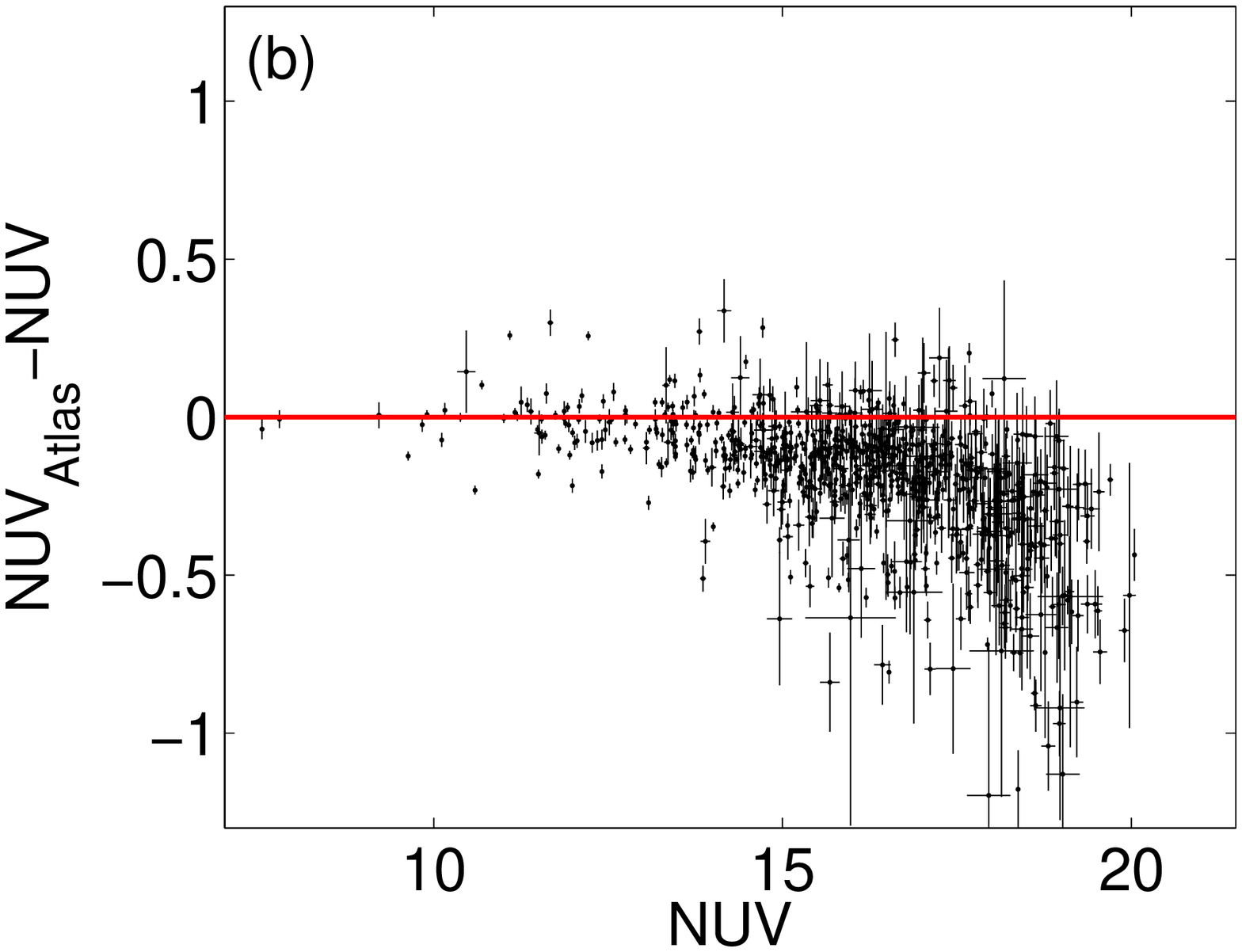}
\includegraphics[width=45mm]{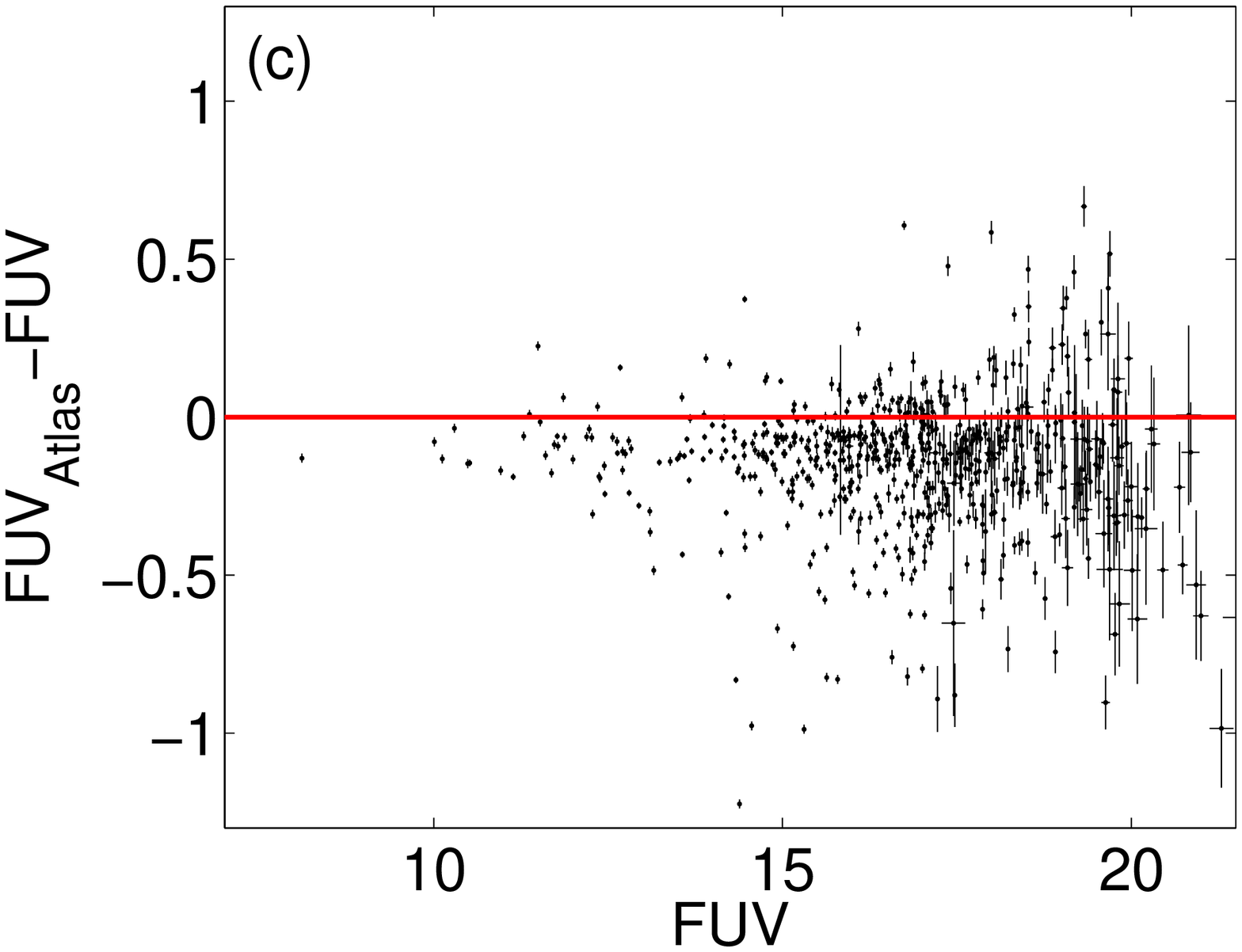}
\includegraphics[width=45mm]{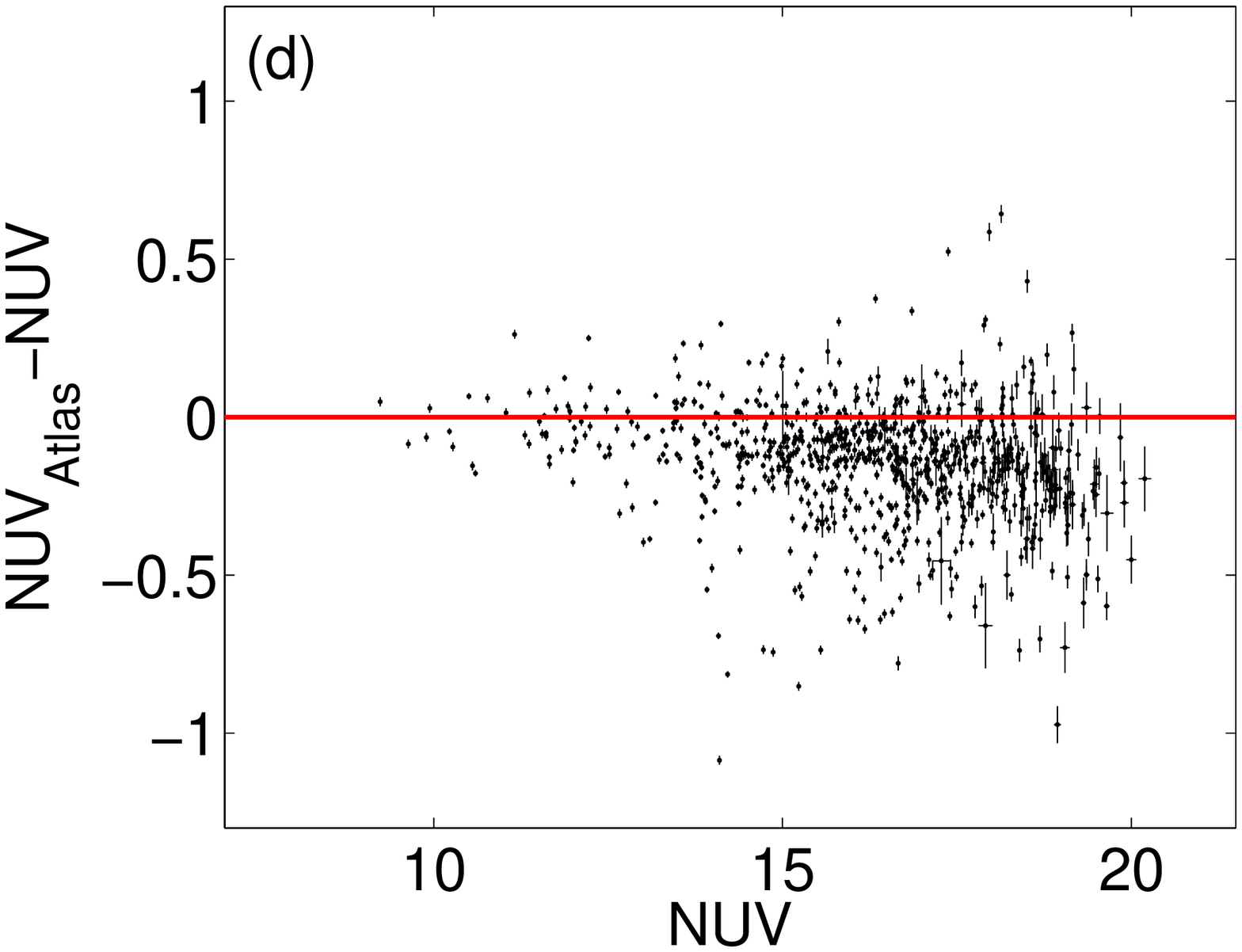}
\includegraphics[width=45mm]{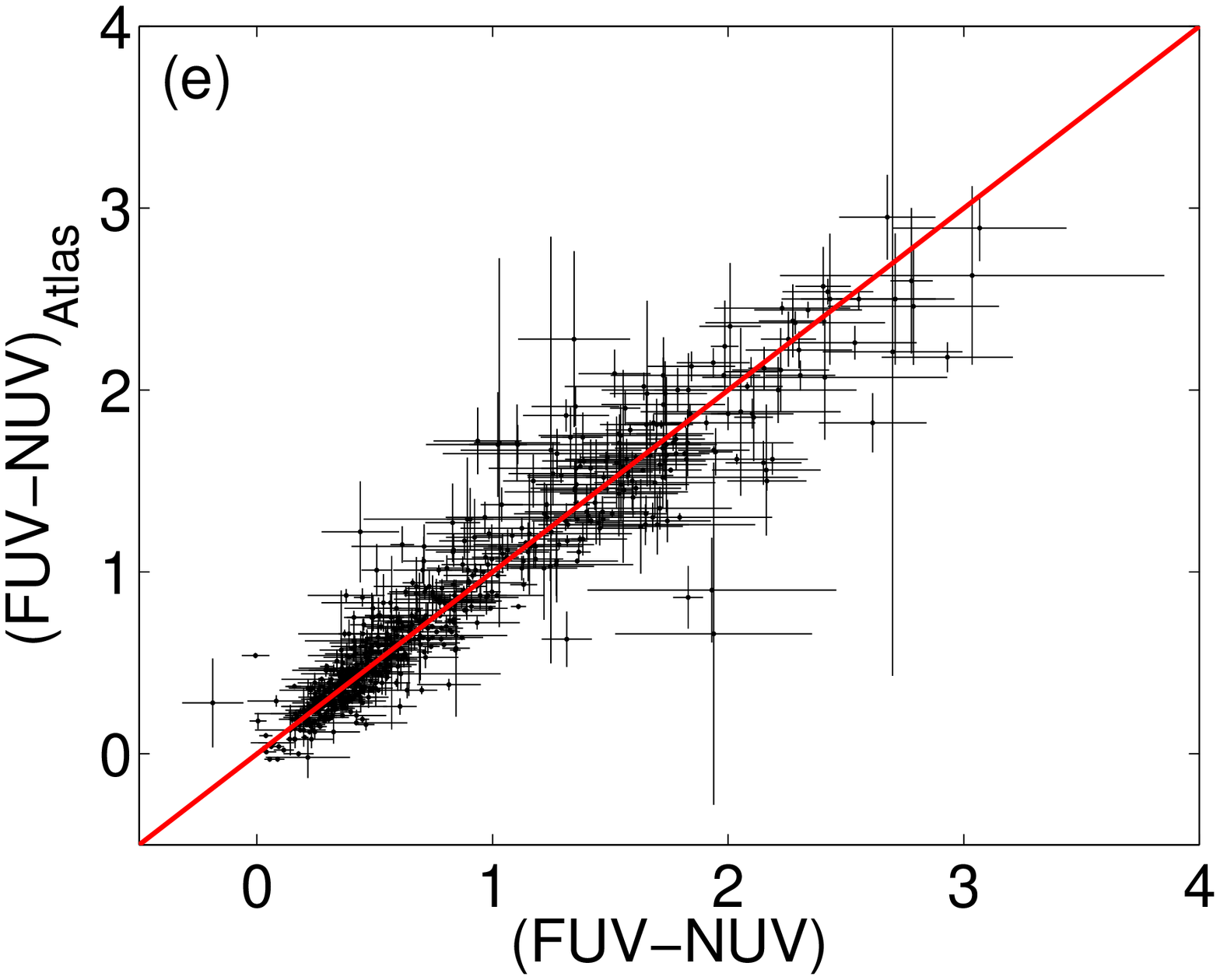}
\includegraphics[width=45mm]{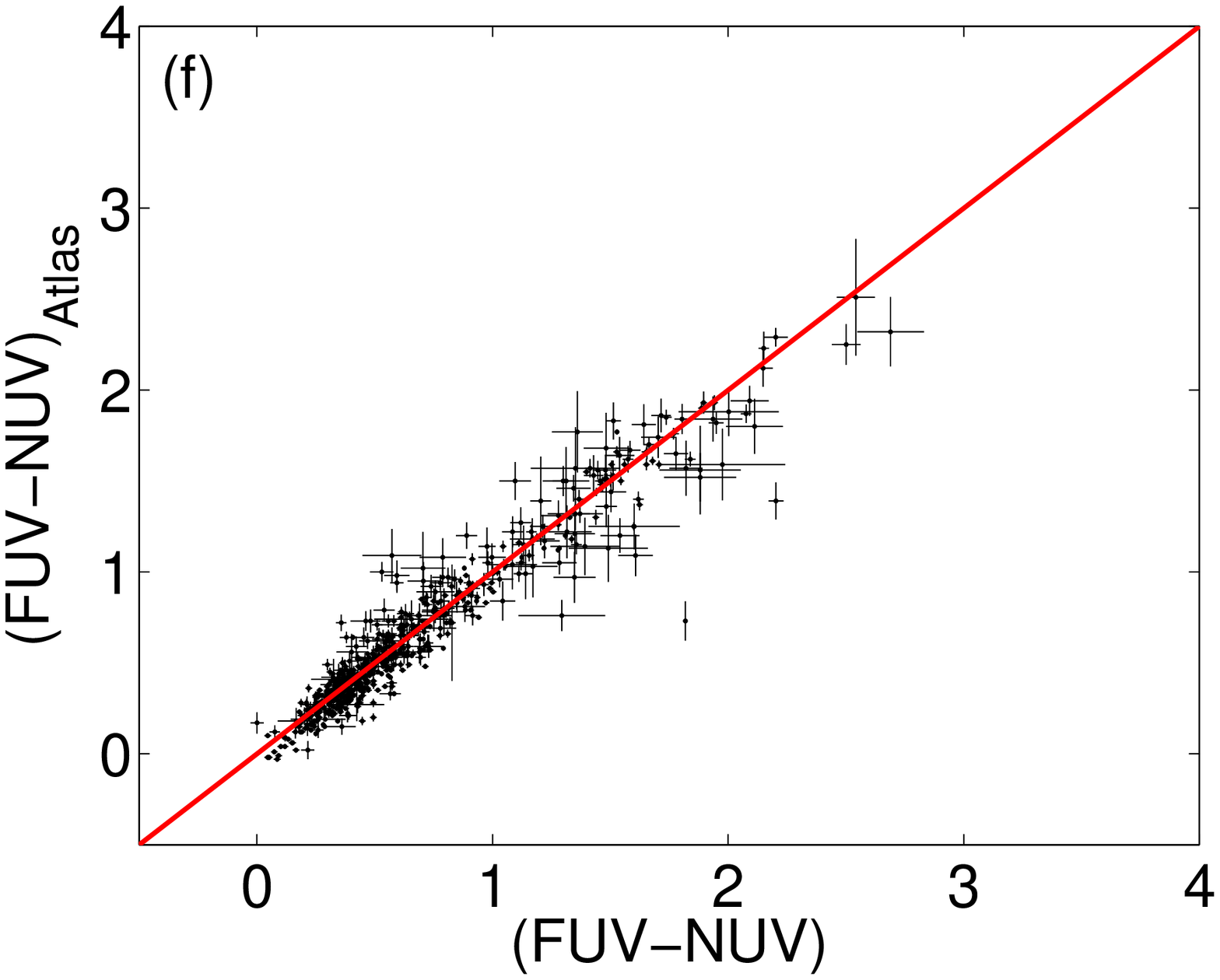}
\includegraphics[width=45mm]{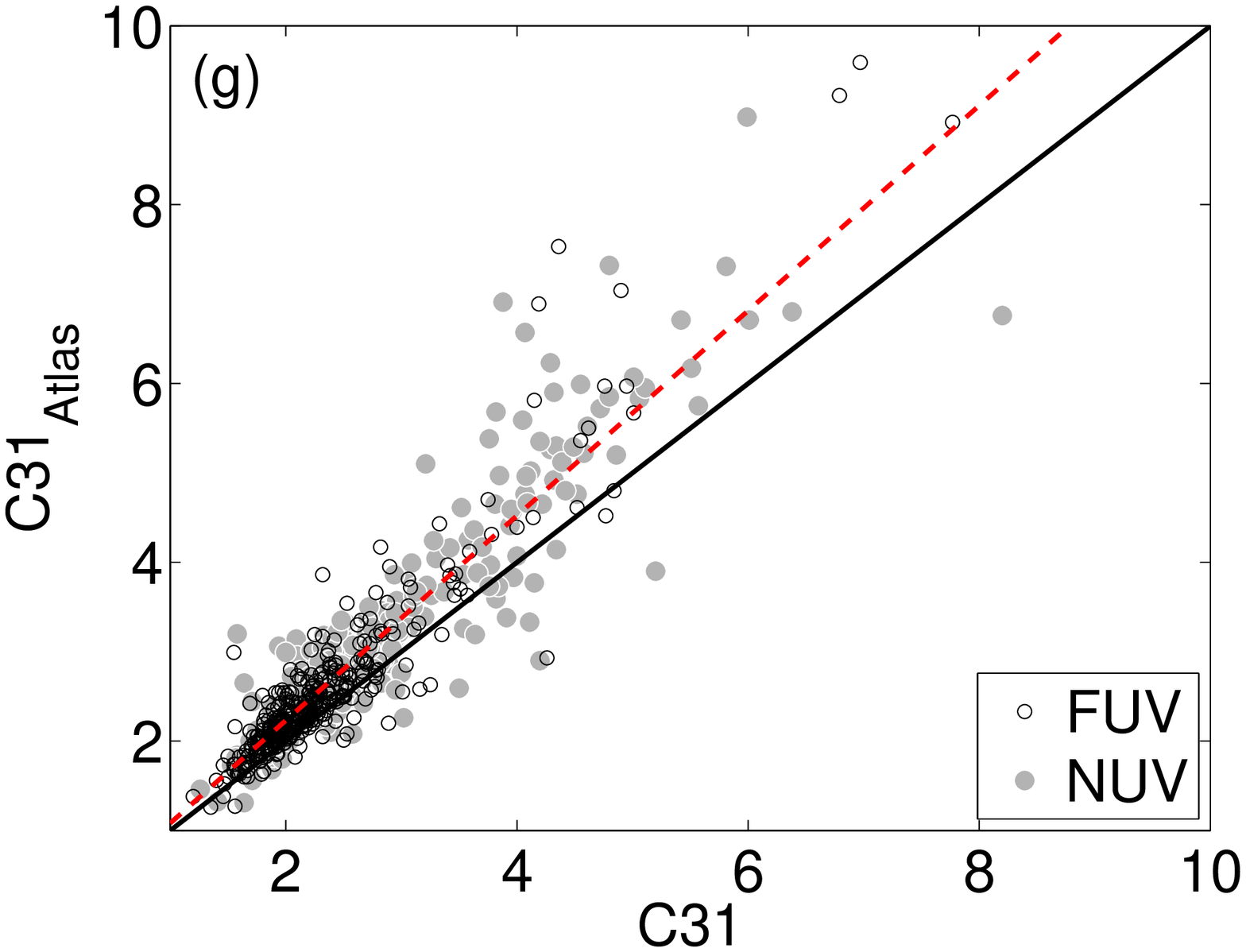}
\includegraphics[width=45mm]{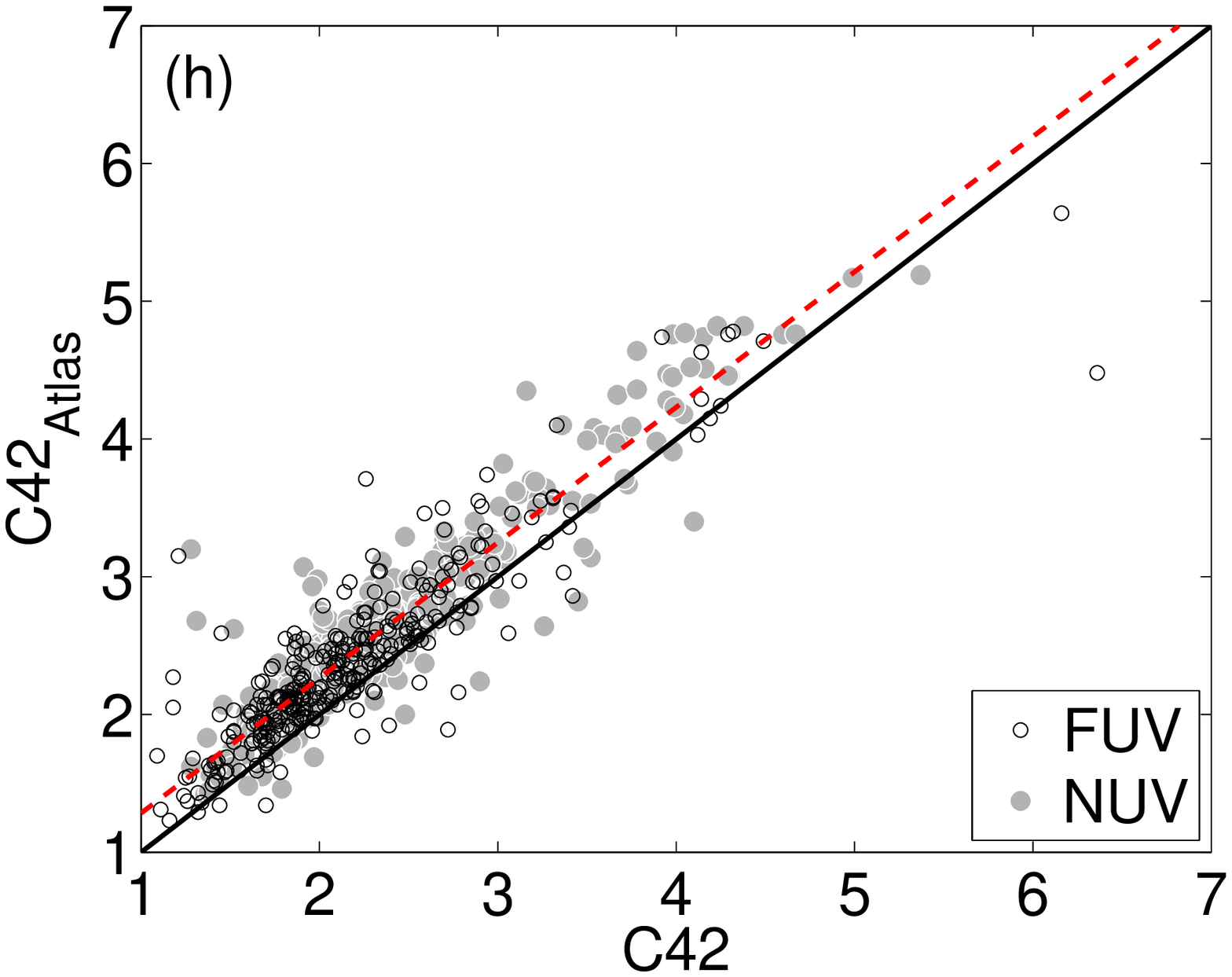}
\caption{Photometric comparisons between the the \textit{GALEX} Atlas and our catalog.
\textit{(a)--(b)} FUV and NUV asymptotic magnitudes difference between \textit{GALEX} Atlas and our catalogs vs. asymptotic magnitudes in \textit{GALEX} Atlas.
\textit{(c)--(d)} FUV and NUV aperture magnitudes difference vs aperture magnitudes in \textit{GALEX} Atlas.
\textit{(e)} Asymptotic (FUV $-$ NUV) in our catalog vs. that in \textit{GALEX} Altas.
\textit{(f)} Aperture (FUV $-$ NUV) in our catalog vs. that in \textit{GALEX} Altas.
\textit{(g)--(h)} our C31 and C42 vs. those in \textit{GALEX} Atlas, and the red dashed lines stand for the linear fittings of the galaxies in the NUV.
\label{fig4}}

\end{figure}

We show the photometric comparisons between \textit{GALEX} Atlas and our sample in Figure \ref{fig4}.
Figure \ref{fig4}a--d display the differences of asymptotic and aperture magnitudes between our catalog
and \textit{GALEX} Atlas. There are 94\% and 93\% galaxies in the FUV and NUV  with the difference of
asymptotic magnitudes within 0.5 mag. We check the galaxies with differences larger than 0.5 mag and
find that these differences might be caused by the different sky-background estimation and elliptical
parameters inherited from their parent catalogues. We fit a sky
background map for each galaxy, while \textit{GALEX} Atlas only use a single background value.
For example, the average background of PGC 4085 (ESO 243-G041) given by \textit{GALEX} Atlas
is 5.59 $\times$ 10$^{-4}$ counts s$^{-1}$ in the NUV, while the average value of our computed background
is 2.24 $\times$ 10$^{-3}$ counts s$^{-1}$, which leads to the magnitude difference of 0.59.
After checking the intensity image, our fitted background is more reasonable, since
the flux of no-signal area around the galaxies is about 2 $\times$ 10$^{-3}$ counts s$^{-1}$.
There are many galaxies in \textit{GALEX} Atlas and our sample with different axis ratios and D25 diameters.
When galaxies become fainter, the difference of the axis ratios can reach a factor of 9,
which leads to different growth curves and results in different magnitudes.
For example, the axis ratio of PGC 4190 (NGC 0407) given by HyperLeda is 6.9,
while \textit{GALEX} Atlas gives 4.3. 
These differences lead to the different magnitudes between \textit{GALEX} Atlas and our sample.

Figure \ref{fig4}e and f show the differences between the (FUV-NUV) colors of
asymptotic and aperture magnitudes. The colors of our sample is consistent with
those in the catalogues of \textit{GALEX} Atlas with standard deviations of 0.19 and
0.13 for asymptotic and aperture colors, respectively.

The differences of the concentration indices for the two samples are displayed in Figure \ref{fig4}g and h.
The concentration indices in both FUV and NUV of our samples are on average smaller than those in \textit{GALEX}
Atlas, probably due to the differences of the growth curves.

\section{Analyses and Some Applications} \label{aa}
\subsection{UV Properties of the galaxies}
\begin{figure}
\includegraphics[width=45mm]{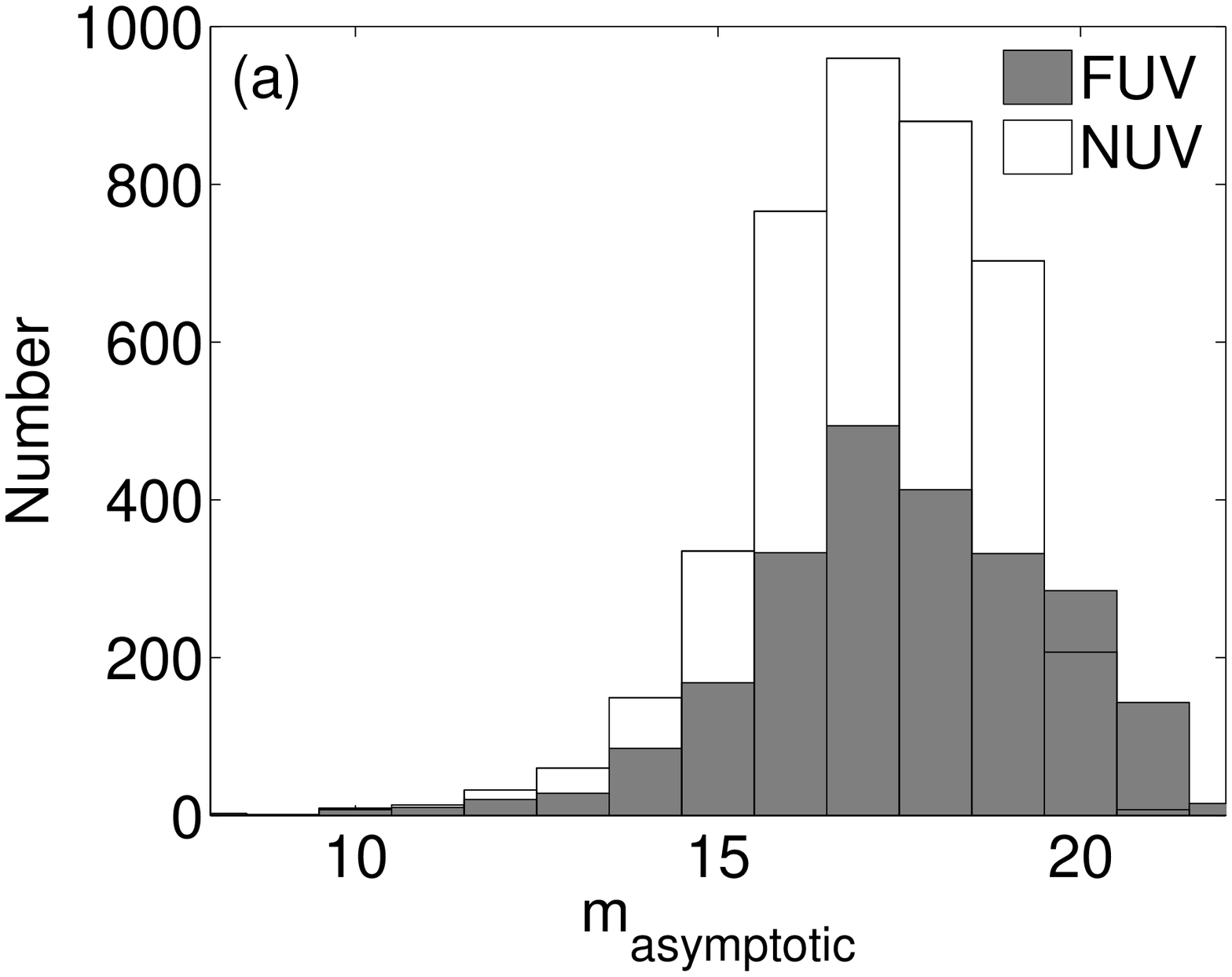}
\includegraphics[width=45mm]{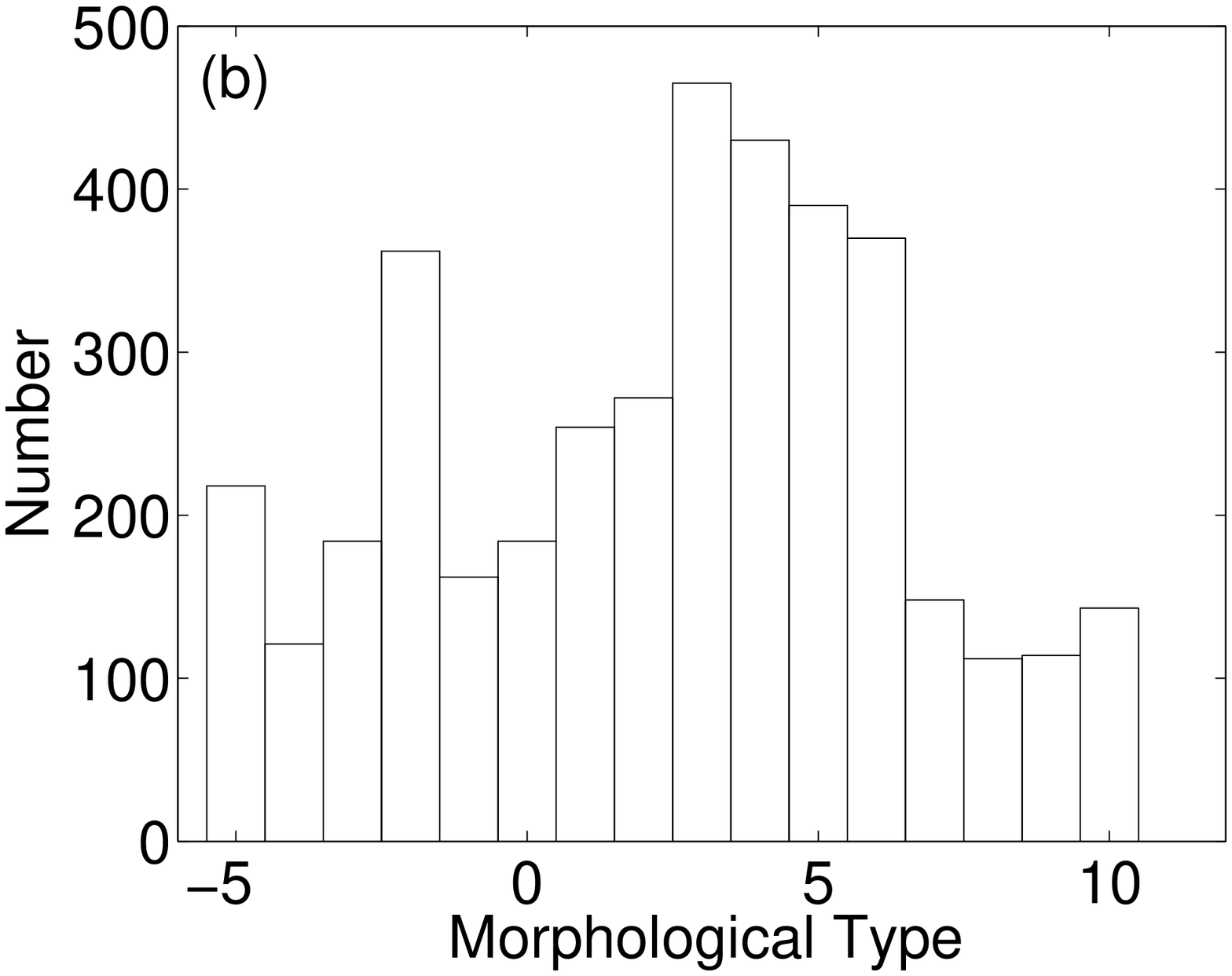}
\includegraphics[width=45mm]{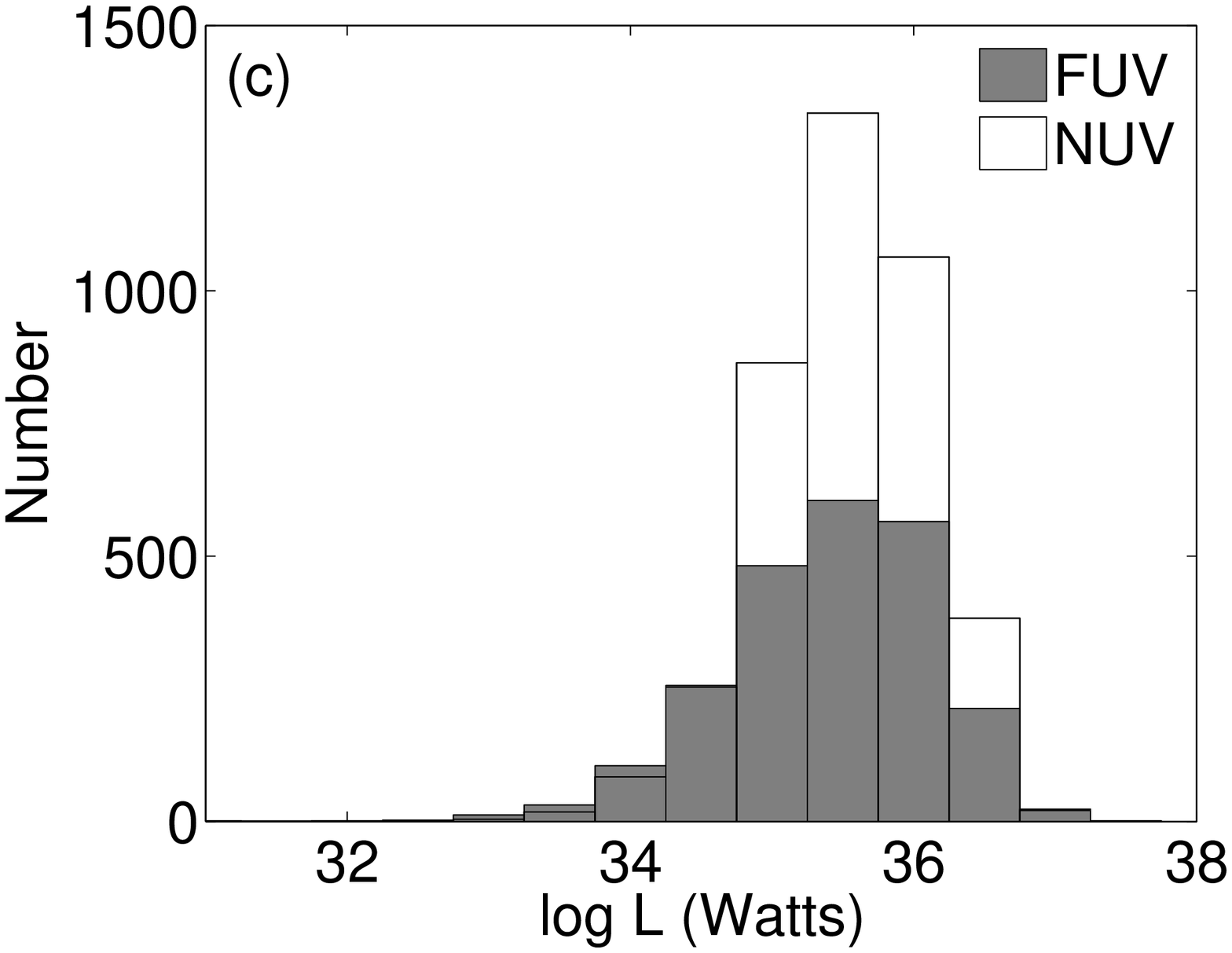}
\includegraphics[width=45mm]{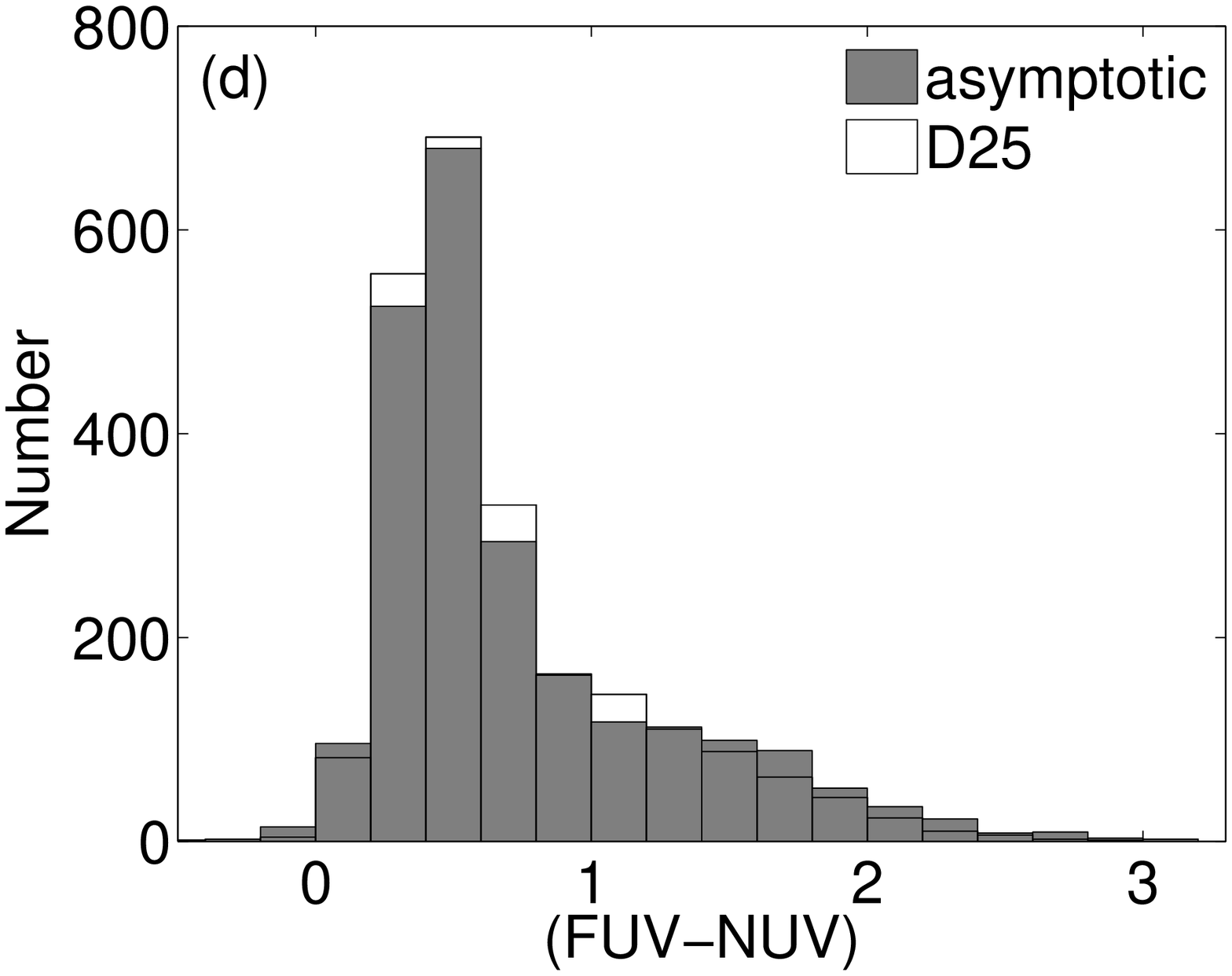}
\includegraphics[width=45mm]{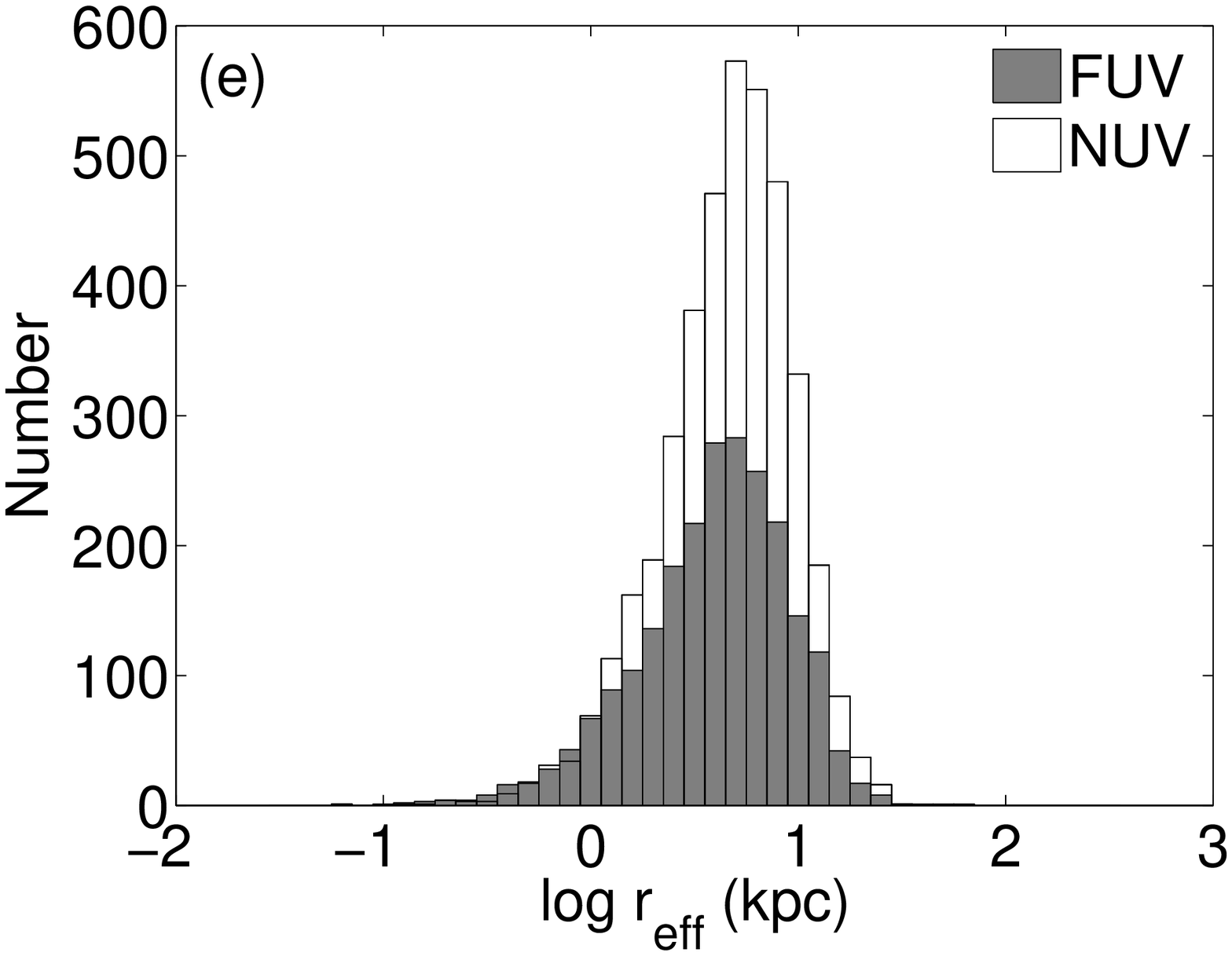}
\includegraphics[width=45mm]{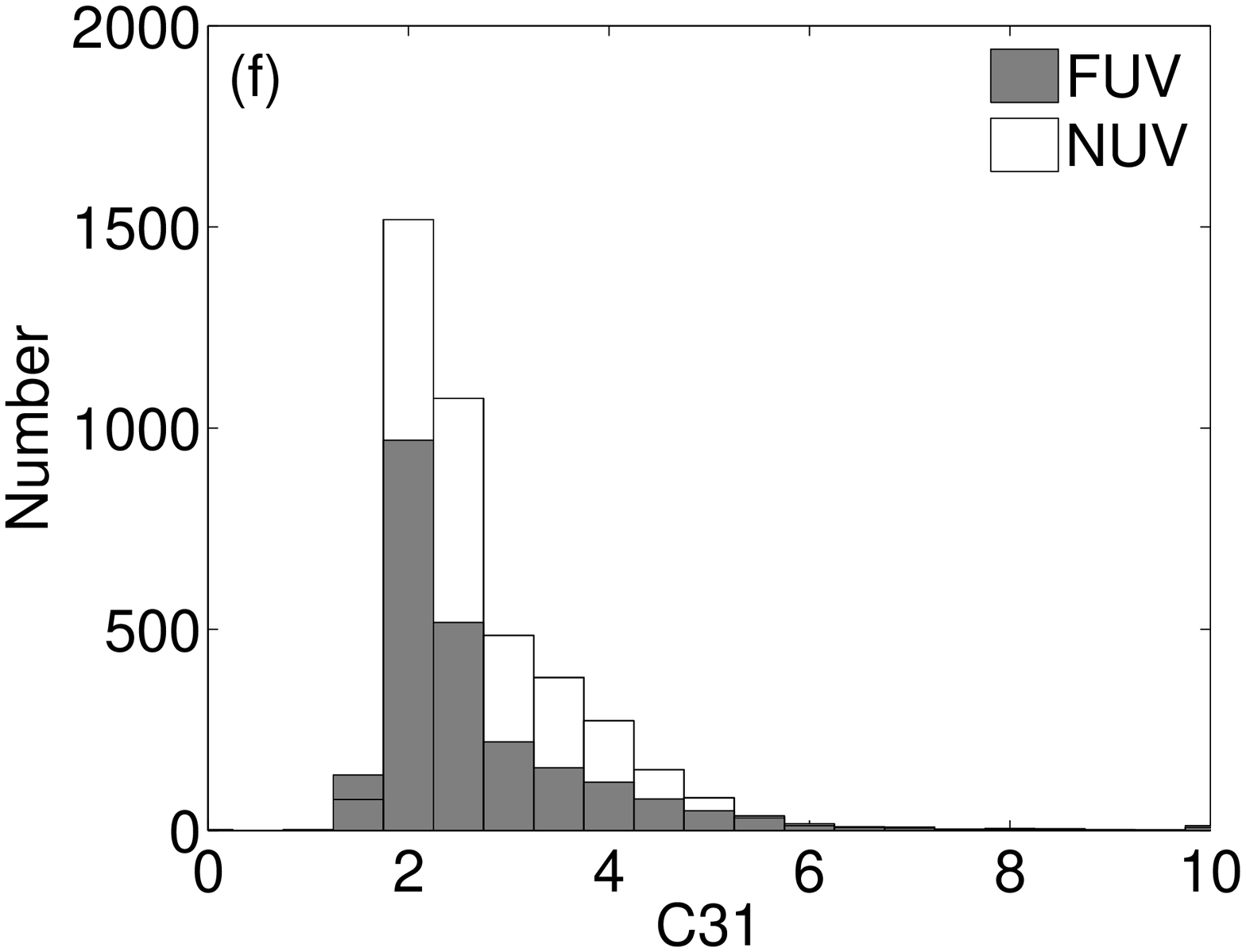}
\includegraphics[width=45mm]{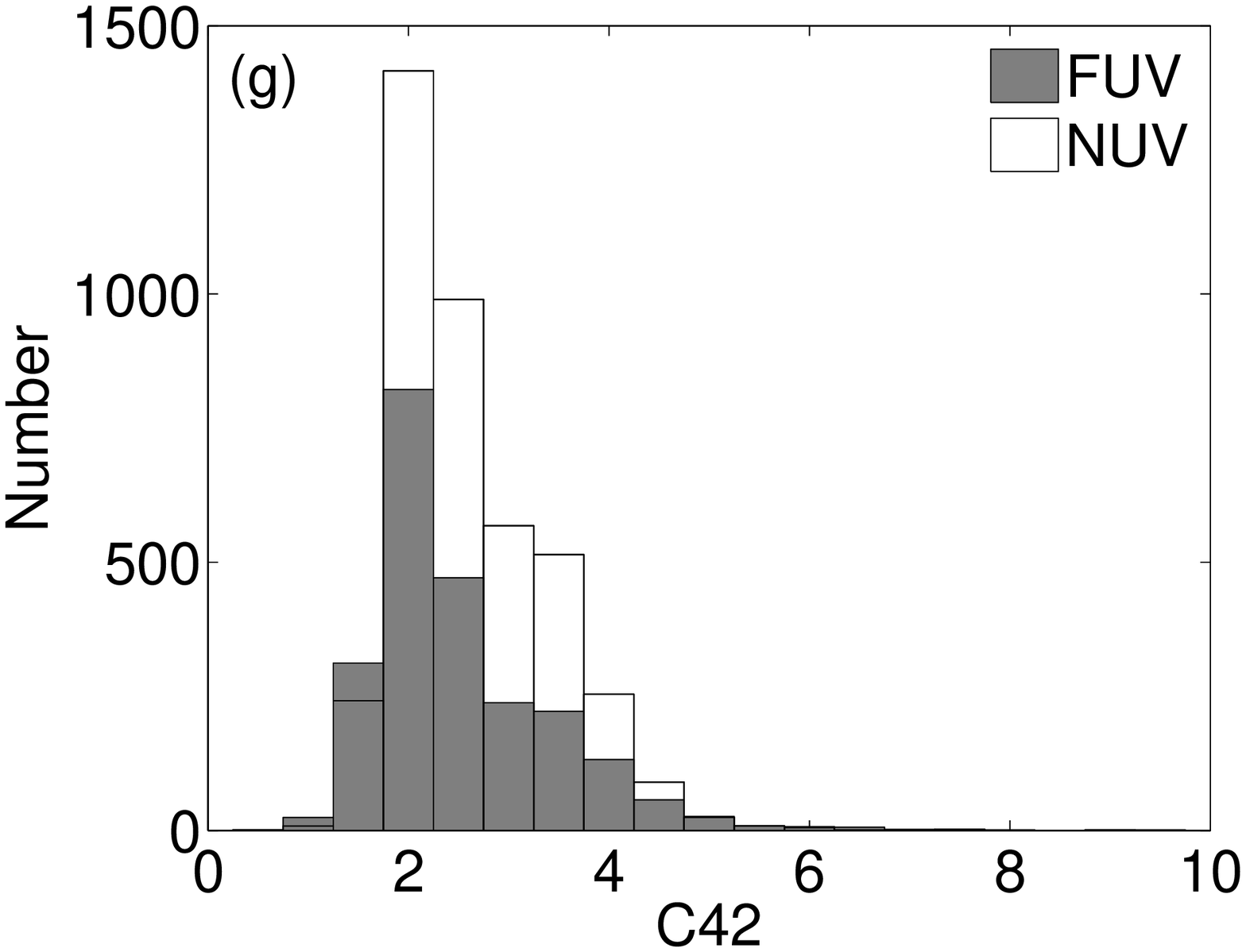}
\includegraphics[width=45mm]{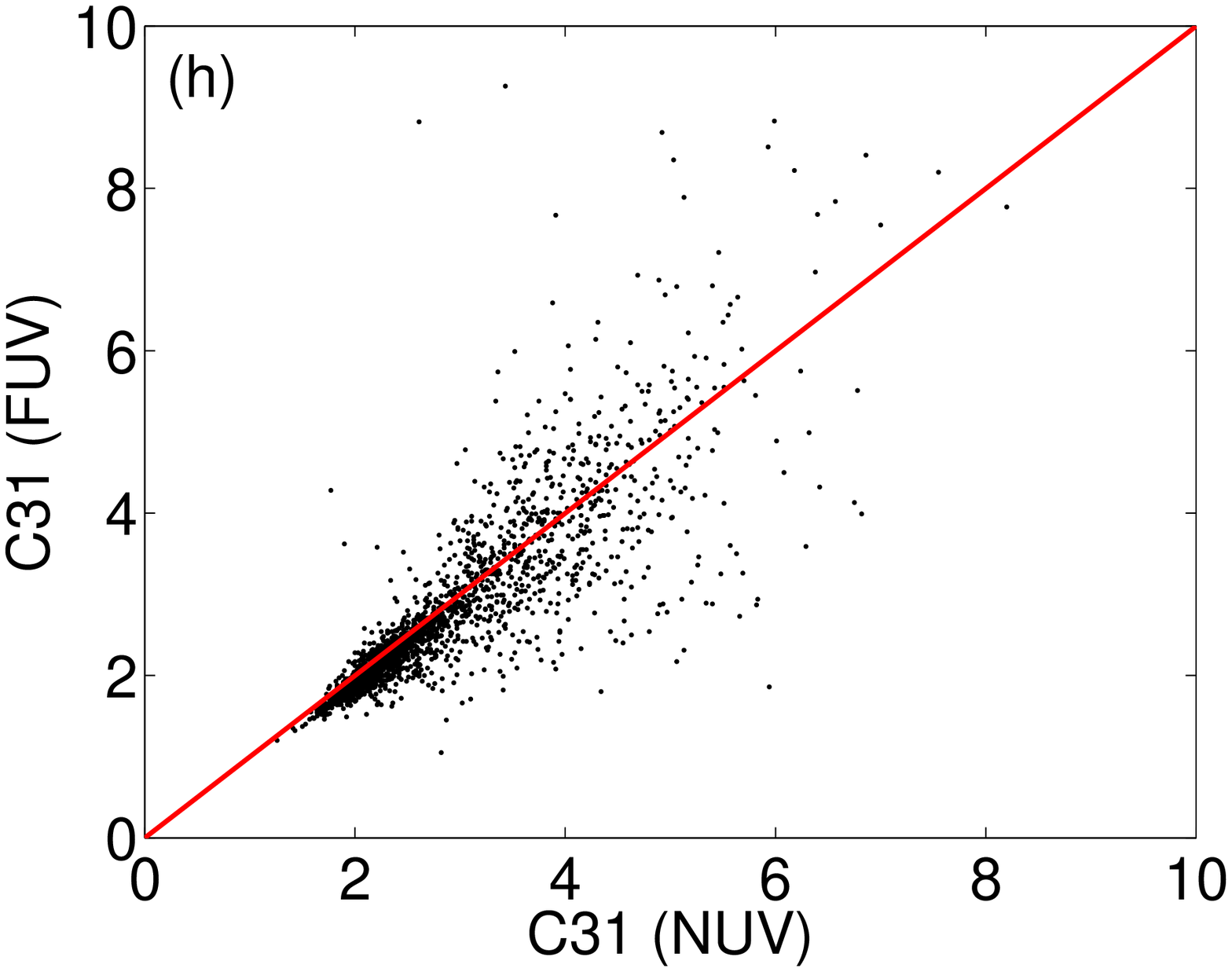}
\centering
\includegraphics[width=45mm]{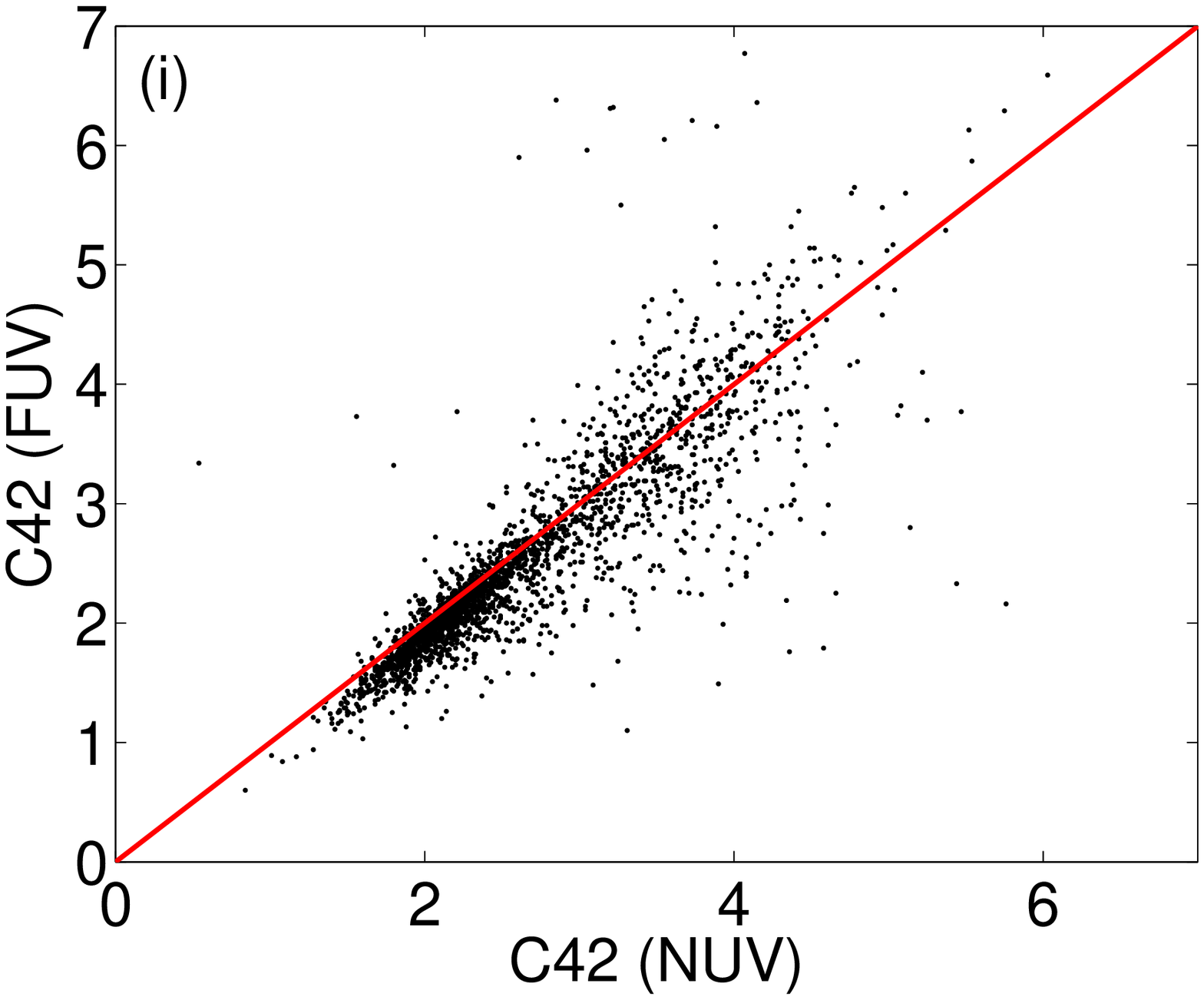}
\includegraphics[width=45mm]{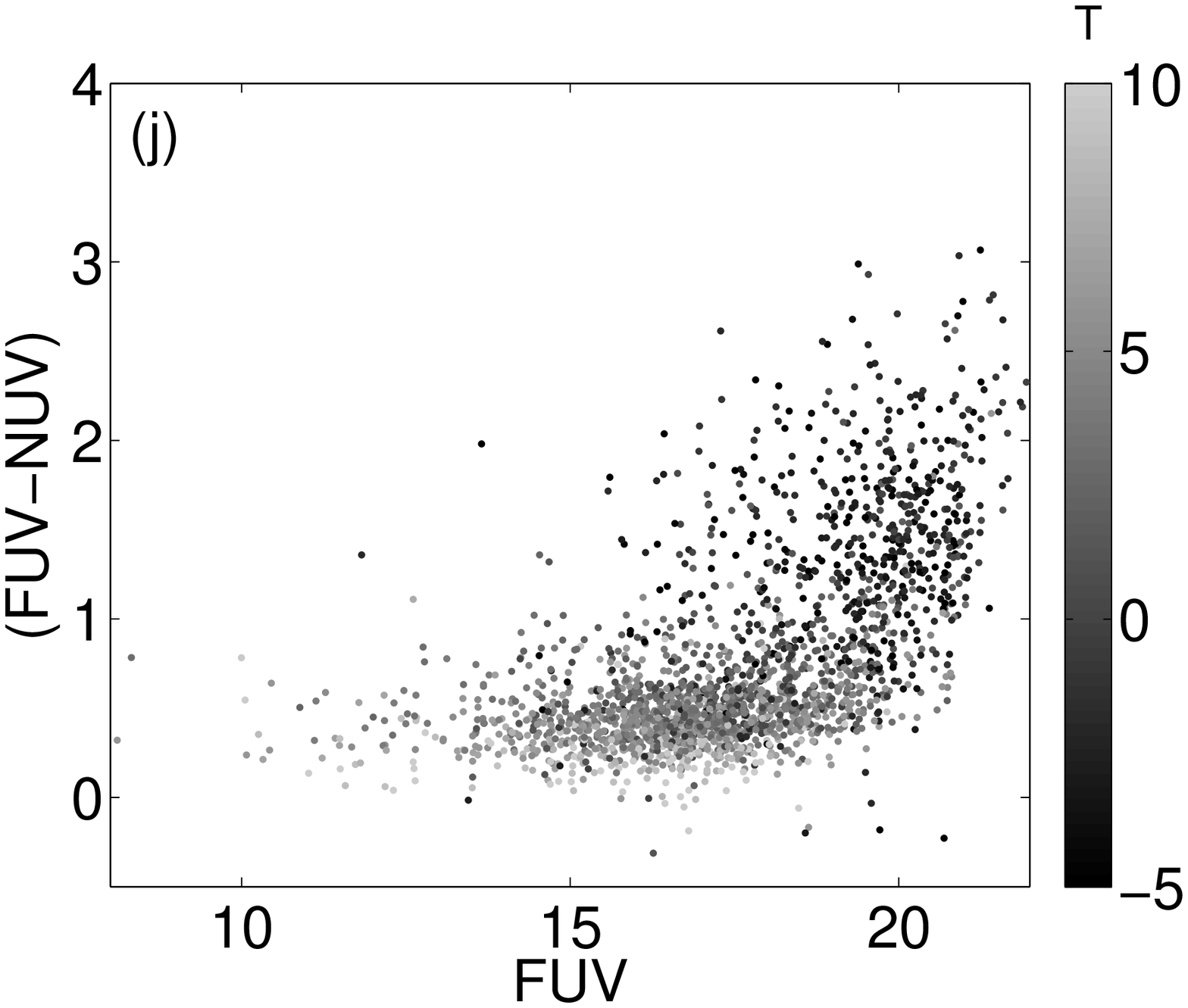}
\caption{Statistical UV properties of our galaxy samples.
\textit{(a)} Distribution of asymptotic magnitudes in the FUV (gray-shade) and NUV (solid).
\textit{(b)} Distribution of the morphological types.
\textit{(c)} Luminosity in watts.
\textit{(d)} (FUV-NUV) color.
\textit{(e)} Effective radius in kpc.
\textit{(f)--(g)} C31 and C42 concentration indices.
\textit{(h)--(i)} C31 and C42 in FUV vs. those in NUV.
\textit{(j)} Asymptotic (FUV-NUV) color vs. asymptotic FUV  magnitudes.
             The point color reflects the morphology type as shown by the color bar to the right.
}
\label{fig5}
\end{figure}

Figure \ref{fig5} shows the distributions of UV properties of the galaxies
in our catalog. Figure \ref{fig5}a-g presents the histograms of asymptotic magnitudes,
morphological type, luminosities, (FUV $-$ NUV) colors, effective radii,
and concentration indices. The median asymptotic magnitudes in FUV and NUV are
17.55 and 17.20 mag, respectively.
In our sample, about 5$\%$ galaxies show asymptotic magnitudes larger
than D25 magnitudes. For these galaxies, the diameters corresponding to
asymptotic magnitudes are smaller than the D25 diameters due to the lack of
diffuse emission. The residual flux after background subtraction within
D25 ellipses leads to additional UV flux in the D25 magnitudes.

There are 1030 (25\%) and 2899 (70\%) early and late-type galaxies (Figure \ref{fig5}b),
and 209 (5\%) galaxies without morphological types given by HyperLeda. 
Figure \ref{fig5}c presents the luminosity distributions . The median FUV
and NUV luminosities are 7.8 $\times$ 10$^8$ and 9.1 $\times$ 10$^8$ $L_{\odot}$
(3.0 $\times$ 10$^{35}$ and 3.5 $\times$ 10$^{35}$ W), respectively.
Ten galaxies in our catalog are ultraviolet-luminous galaxies (UVLGs; \citealt{Heckman05})
defined as the FUV luminosity larger than 2$\times$10$^{10}$ $L_{\odot}$ ($7.6\times10^{36}$ W),
which are extremely rare in the local universe. Among the ten galaxies, PGC 2248, 36466,
59214 have been identified by \textit{GALEX} Atlas. PGC 53898 and 71035 are Seyfert
galaxies \citep{Veron06}. PGC 4007 and 17625 are a luminous infrared galaxies (LIRG; \citealt{Haan11,Sanders03}).
The other three galaxies, PGC 23064, 30400, and 51865, are star-forming
spiral galaxies with very UV-luminous arms. PGC 70348 (NGC 7469) is defined as a UVLG in \textit{GALEX} Atlas but not
qualified in our sample. Our FUV luminosity of this galaxy is about 7.3$\times$10$^{36}$ W
, close to the threshold.

Figure \ref{fig5}d presents the color distributions of asymptotic and D25 magnitudes.
The majority of galaxies has color bluer than (FUV $-$ NUV) = 1, which has been used to separate
foreground stars. The red tail of the distributions is populated by early-type galaxies.
The distributions of effective radii in kiloparsecs are shown in Figure \ref{fig5}e.
The median effective radii for FUV and NUV are 4.3 and 5.0 kpc, respectively.
The distributions and comparisons of concentration indices are presented in
Figure \ref{fig5}f-i. The galaxies are on average slightly
more concentrated in the NUV than in the FUV by differences of 0.03 in C31 and
0.10 in C42,
which may be due to the different fractions of the bulge component
in spiral galaxies in these two bands \citep{Gil07}. Figure \ref{fig5}j shows the color
as a function of the FUV magnitude. The late-type galaxies are brighter and located in a
narrow color range, while the early-type galaxies are fainter and distributed with
broader and redder color span.

\subsection{Effective Surface Brightness}
\begin{figure}
   \centering
   \includegraphics[width=12cm]{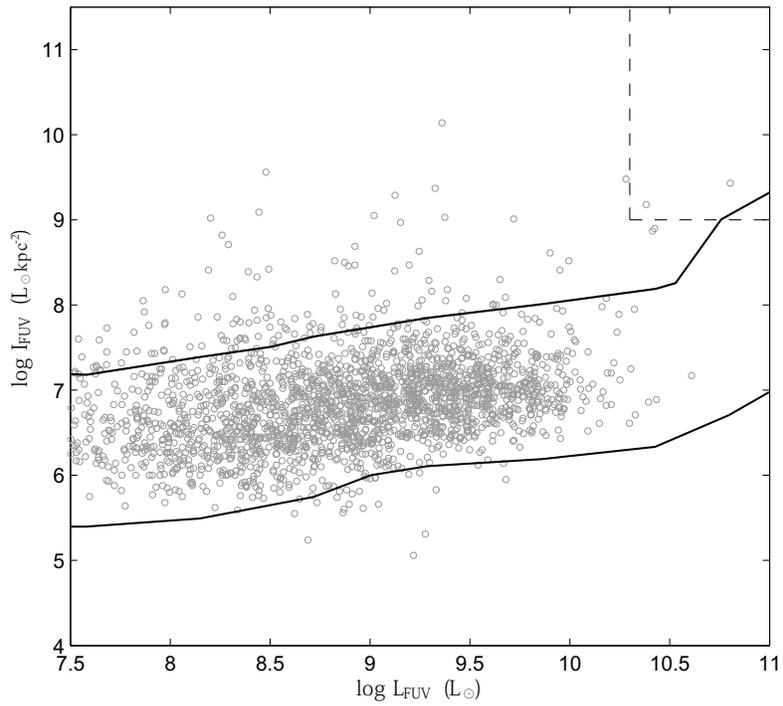}
   \caption{FUV surface brightness vs. FUV luminosity. UV properties for the galaxies
   of our sample are shown as gray empty circles. \textbf{The contours stand for
   the GR1/DR3 sample with FUV detections, which are defined as
   enclosing 84\% of galaxies in the GR1/DR3 sample \citep{Hoopes07}. }
   The dashed line shows the region typically populated by LBGs.\label{fig6}}
\end{figure}

The plot of effective surface brightness versus the
FUV luminosity is presented in Figure \ref{fig6}. This figure shows a trend of slight increasing
effective surface brightness with increasing luminosity, indicating that
galaxies with large effective radii turn to be more luminous \citep{Hoopes07}.
\textbf{The effective surface brightness distribution of our sample is more concentrated
than \citet{Hoopes07} sample, since the contours in Figure \ref{fig6} enclose 96\% 
of the galaxies in our sample but only 84\% in their sample. }

The compact UVLGs are UVLGs defined with a surface brightness $I_{FUV} >$ 10$^{9}$ $L_{\odot}$ kpc$^{-2}$
\citep{Hoopes07}. They have characteristics that are remarkably similar to the Lyman break galaxies
(LBGs), such as SFRs, metallicities, morphologies, kinematics, and attenuations
\citep{Hoopes07,Overzier08,Basu-Zych07}.
Only 12 galaxies in our sample have the surface brightness $I_\mathrm{FUV} >$ 10$^{9}$ $L_{\odot}$ kpc$^{-2}$.
Two of them, PGC 59214 and PGC 71035, have both high luminosity
and high surface brightness, which can be classified as compact UVLGs.
PGC 59214 is a BL Lac object and PGC 71035 is a Seyfert 1.5 \citep{Veron06}.
Both of them have very compact nuclei in the FUV, and the optically thick accretion
disks around super-mass black holes probably dominates their FUV luminosity
\citep{Malkan82,Ward87,Sanders89}.

\subsection{(FUV $- K$) tracing galactic morphology}
Compared to the optical colors \citep{Weiner05,Pozzetti10,Talia14}, the combination
of UV and IR should be more efficient in separating the early and late-type galaxies,
given their different SFRs and dust attenuation \citep{Gil07}.
Since the UV emission is very sensitive to the presence of recent star
formation activity and the $K$ band emission is sensitive to the accumulated star
formation \citep{Munoz07}, the (FUV $- K$) color provides a robust discrimination
between early and late-type galaxies as analyzed by \citet{Gil07}.

\begin{figure}
   \centering
   \includegraphics[width=62mm]{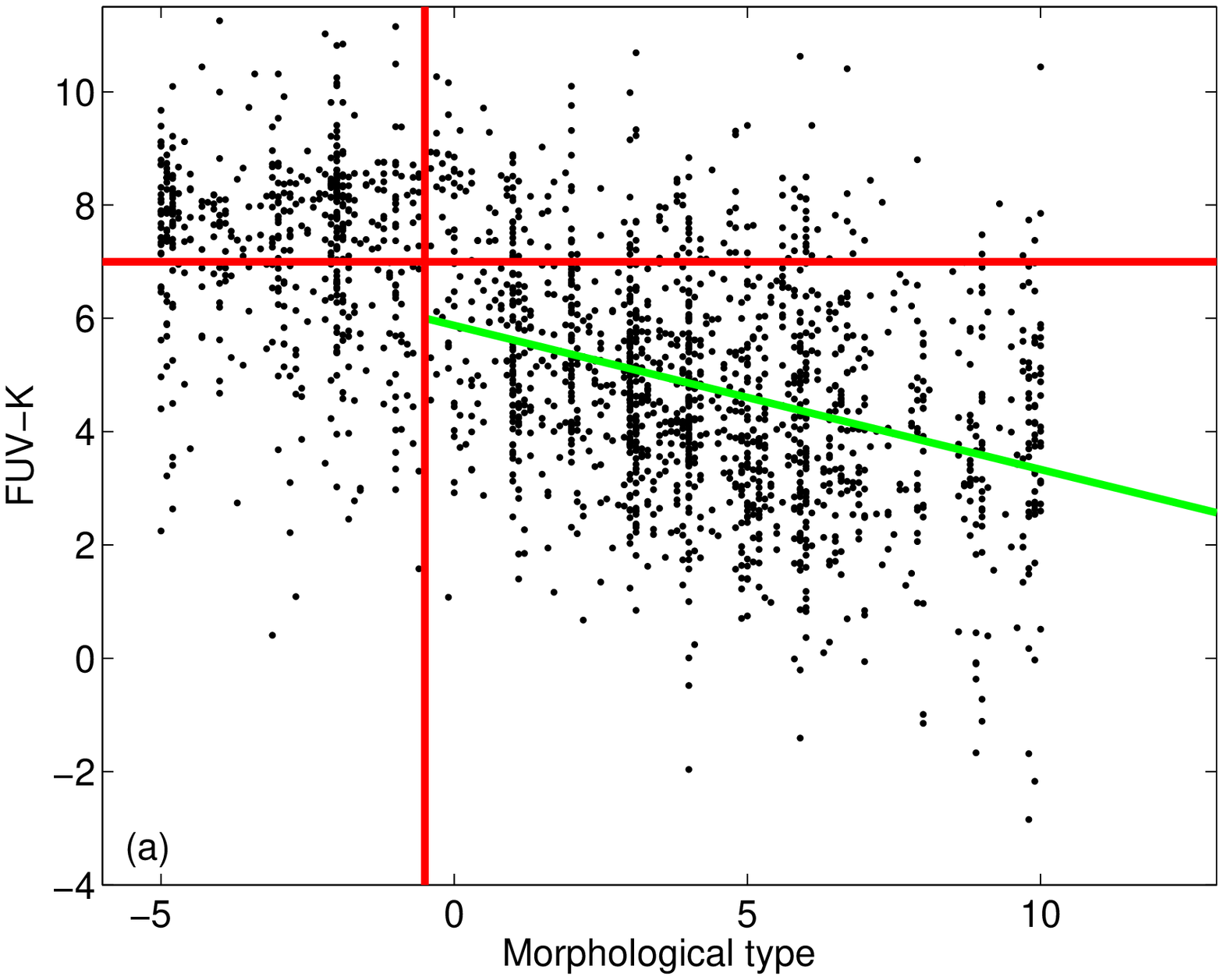}
   \includegraphics[width=62mm]{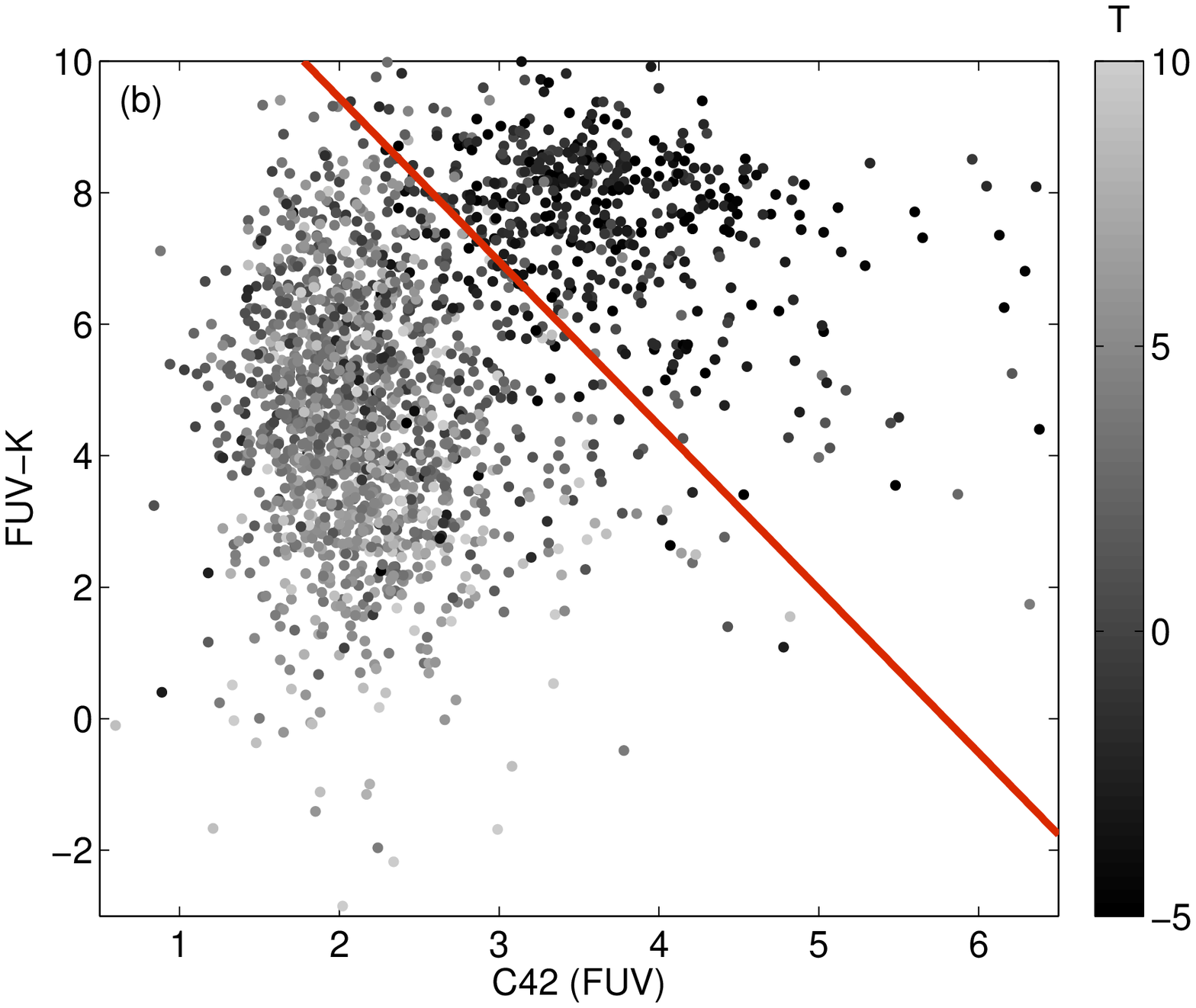}
   \caption{
   \textit{(a)} (FUV - $K$) vs. galactic morphological type. The red vertical line
   shows the separation between early and late-type galaxies. The red horizontal
   line is the demarcation point of (FUV - $K$) that we use to separate these two
   types of galaxies. The green line give the linear fit to the relation between
   (FUV - $K$)  and morphological type for late-type galaxies.
   \textit{(b)} (FUV$-$K) vs. the FUV concentration index C42.
   The point color reflects the morphology type as shown by the color bar to the
   right, and the red line best separates early and late-type galaxies. 
    }
   \label{fig7}
\end{figure}

Figure \ref{fig7}a shows the relation between (FUV $- K$) color and galactic morphology.
Here we convert the 2MASS $K$ band Vega magnitudes to the AB system by adding 1.84 mag \citep{Munoz09,Cohen03}.
We calculate the dividing (FUV $- K$) color to separate early and late-type galaxies
by minimizing the number of galaxies that are misclassified by the dividing magnitude,
which leads to (FUV $- K$) = 6.84. About 68\% early-type galaxies and 86\% late-type
galaxies are correctly classified. The (FUV $- K$) is linearly correlated to the morphology of
late-type galaxies. The best linear fit to the relation (green line in Figures \ref{fig7})
is
\begin{equation}
  \mathrm{(FUV }- K) = \mathrm{(5.8\pm0.1) - (0.25\pm0.02) \times T}.
\end{equation}
Here T is the morphological type given by HyperLeda and the errors of the
parameters are derived from the linear fitting.

The concentration indices have also been used as a classification tool
\citep{Abraham96,Bershady00}. The (FUV$ - K$) color together with concentration indices are expected
to show better separation of different type galaxies, which is discussed in \citet{Gil07}.
We present the FUV concentration index C42 as a function of the  (FUV $ - K$) color in
Figure \ref{fig7}b. Early-type galaxies have larger C42 and redder (FUV $ - K$). The best
separation line in the plane of this figure is computed as
\begin{equation}
	(\mathrm{FUV }- K) = 14.4\pm0.3 - (2.5\pm0.1) \times \mathrm{C42_\mathrm{FUV}}.
  \end{equation}
With this, 76\% and 94\% of early and late-type galaxies are correctly classified.

\subsection{IRX-$\beta$ relation}
\label{damr}
The IR excess (IRX) is widely used as a good tracer of the dust attenuation in galaxies,
and the slope of the UV continuum ($\beta$) is considered as a proxy to estimate the IRX when
IR data are not available, which is known as IRX-$\beta$ relation \citep{Daddi07,Reddy09,Takeuchi10}.
IRX is defined as log($f_\mathrm{TIR}/f_\mathrm{FUV}$), where $f_\mathrm{TIR}$ is the total infrared flux,
and $\beta$ can be estimated by (FUV $-$ NUV).
Here, the IRX-$\beta$ relation is not valid for early-type galaxies, since a substantial
part of dust heating is due to photons emitted by old stars which do not emit primarily
in UV and the dust attenuation would be overestimated by IRX \citep{Buat11}.

\begin{figure}
   \centering
   \includegraphics[width=\linewidth]{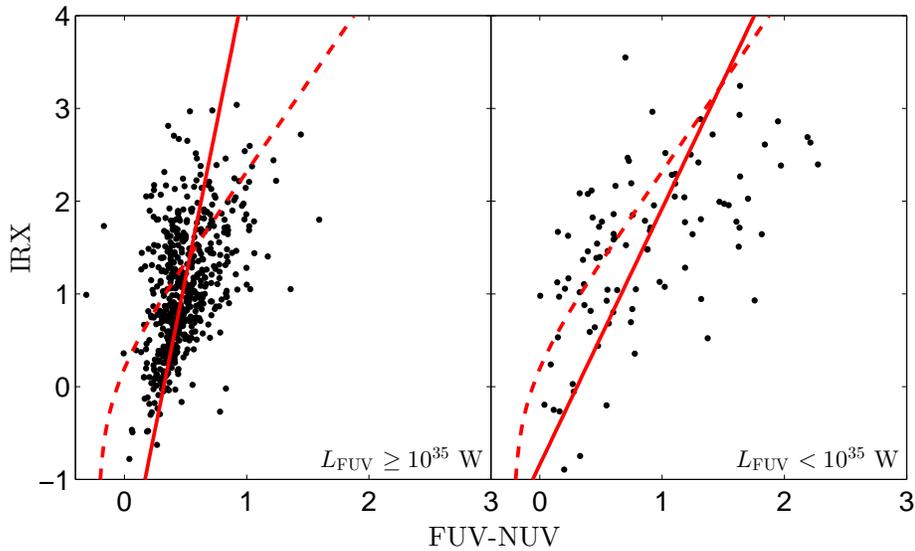}
   \caption{IRX-$\beta$ relations for galaxies with $L_\mathrm{FUV} \geq 10^{35}$ W (left) and galaxies with
	   $L_{\mathrm{FUV}} < 10^{35}$ W (right). The dashed curves show the IRX-$\beta$ relation of starburst
	   galaxies derived by \citet{Meurer99}. The solid lines are the robust linear fits to the relations of our
	   samples.
     \label{fig8}}
\end{figure}

Figure \ref{fig8} shows the (FUV $-$ NUV) color against IRX for our late-type galaxies together
with the local starburst relation derived from \citet{Meurer99}. Here we define
$f_\mathrm{FUV} = \lambda f_{\lambda}$ and adopt the formula proposed by \citet{Sanders96} to derive
the $f_\mathrm{TIR}$, which is estimated using the 12, 25, 60 and 100 $\mu$m from IRAS data. We separate
our galaxies into two groups according to their FUV luminosities.  In Figure \ref{fig8}, we can see
that most of our galaxies are located below the relation of \citet{Meurer99}. The galaxies with brighter
FUV luminosities ($L_{\mathrm{FUV}} \geq 10^{35} W$) have a tigher relation than that of fainter galaxies.
Both groups have steeper IRX$-\beta$ relations than that of \citet{Meurer99}. We linearly fit the relations
of these two groups, which are formulated as
\begin{equation}
	\mathrm{IRX} = (6.54\pm0.01) \times (\mathrm{FUV} - \mathrm{NUV}) - (2.10\pm0.01), L_\mathrm{FUV} \geq 10^{35} \mathrm{W},
\end{equation}
\begin{equation}
	\mathrm{IRX} = (2.76\pm0.05) \times (\mathrm{FUV} - \mathrm{NUV}) - (0.84\pm0.07), L_\mathrm{FUV} < 10^{35} \mathrm{W}.
\end{equation}

\section{SUMMARY}
\label{sum}
In this paper, we provide an updated catalog of 4138 nearby galaxies based on the
latest (GR6/GR7) of \textit{GALEX}, which is more than 3 times the number of galaxies
in the original \textit{GALEX} Atlas.
Our samples are selected from the extragalactic catalog of HyperLinda.
The D25 diameter is set to be larger than 1{\arcmin}. Compared with \textit{GALEX} Atlas, we
apply a more precise procedure to estimate the sky background in the \textit{GALEX} images.
Radial FUV and NUV surface brightness profiles are obtained. From these profiles we
calculate asymptotic magnitudes, aperture (D25) magnitudes, UV colors, structural
parameters (effective radii and concentration indices), luminosities, and effective
surface brightness. We also augment our data set with archival optical and infrared
photometry from HyperLeda, 2MASS and IRAS. With this updated catalog, we confirm
that the (FUV $- K$) color provides a good criterion to distinguish early and late-type galaxies,
which can be improved further with the concentration indices.
The IRX-$\beta$ relation is reformulated with our UV-selected nearby galaxies.

The GALEX images and catalogs of our nearby galaxies can be accessed via the website
\url{http://batc.bao.ac.cn/$\sim$zouhu/doku.php?id=projects:galax:start}

\begin{acknowledgements}
The authors acknowledge support from the National Science Foundation of China under
grants NSFC-11273028, NSFC-11333004 and NSFC-11203031, and support from the National
Astronomical Observatories, Chinese Academy of Sciences under the Young Researcher Grant.
We thank Mark Seibert and Scott Fleming for
providing subfunctions to generate false color images. We acknowledge the usage of the
HyperLeda database\footnote{\url{http://leda.univ-lyon1.fr}}.
This research has made use of the NASA/IPAC Extragalactic Database (NED), which is
operated by the Jet Propulsion Laboratory, California Institute of Technology,
under contract with the National Aeronautics and Space Administration.
Some of the data presented in this paper were obtained from the Mikulski Archive
for Space Telescopes (MAST).
This research has made use of the NASA/IPAC Infrared Science Archive,
which is operated by the Jet Propulsion Laboratory, California Institute of Technology,
under contract with the National Aeronautics and Space Administration.
This research made use of Montage, funded by the National Aeronautics and Space
Administration's Earth Science Technology Office, Computation Technologies Project,
under Cooperative Agreement Number NCC5-626 between NASA and the California Institute
of Technology. Montage is maintained by the NASA/IPAC Infrared Science Archive.

\end{acknowledgements}

\clearpage
\pagestyle{empty}


\begin{deluxetable}{rccrrrrr}
\tabletypesize{\scriptsize}
\tablecaption{Our \textit{GALEX} Sample of 4138 Nearby Galaxies \label{Tab1}}
\tablewidth{0pt}
\tablehead{
\colhead{PGC} & \colhead{R.A.}   & \colhead{Dec.}   & \colhead{2$\times$A} & \colhead{2$\times$B}  & \colhead{P.A.} & \colhead{T} & \colhead{Distance}\\
\colhead{}            &  \colhead{(J2000.0)}    & \colhead{(J2000.0)}      &  \colhead{(arcmin)}   & \colhead{(arcmin)}    & \colhead{(deg)} & \colhead{} & \colhead{(Mpc)}\\
\colhead{(1)} & \colhead{(2)} & \colhead{(3)} & \colhead{(4)} & \colhead{(5)} &\colhead{(6)} &\colhead{(7)} &\colhead{(8)}   }
\startdata
      12& 00 00 08.604& $-$06 22 26.00&   1.4&   0.3& 168&   1.3 $\pm$ 1.0&  95   \\
      62& 00 00 46.908& $-$77 34 47.93&   1.0&   0.2&   9&   4.4 $\pm$ 1.9& 153   \\
     120& 00 01 38.316& $+$23 29 00.92&   1.9&   1.1& 160&   5.3 $\pm$ 1.5&  52   \\
     129& 00 01 41.916& $+$23 29 44.95&   1.4&   0.5& 135&   4.8 $\pm$ 1.4&  65   \\
     143& 00 01 58.188& $-$15 27 39.24&  10.5&   3.5&   5&   9.9 $\pm$ 0.3&   0.97\\
     176& 00 02 34.836& $-$03 42 38.92&   1.1&   0.7& 180&   4.0 $\pm$ 0.5&  93   \\
     192& 00 02 48.624& $-$03 36 21.82&   1.1&   0.3&  25&   5.0 $\pm$ 2.0&  90   \\
     215& 00 03 14.184& $-$65 22 11.68&   1.1&   0.5& 167&   4.4 $\pm$ 1.9&  90   \\
     243& 00 03 32.148& $-$10 44 40.81&   1.1&   1.0& ...&$-$2.0 $\pm$ 0.6& 129   \\
     250& 00 03 34.992& $+$23 12 02.92&   1.0&   0.6&  23&   5.5 $\pm$ 0.8& 107   \\
     ...&      ...    &      ...      &   ...&   ...& ...&   ...        & ...   \\
\enddata
\tablecomments{ The parameters are taken from HyperLeda. For galaxies whose distances
are not available, their distances are selected from NED. 
Col.(1): PGC number. Col. (2)--(3): R.A. and Dec. (J2000.0) of the galaxy
center. Right ascensions are in hours, minutes, and seconds, and declinations
are in degrees, arcminutes, and arcseconds. Col. (4): Major-axis diameter of the D25
ellipse. Col. (5): Minor-axis diameter. Col. (6): Position angle ( P.A.). Col. (7):
Morphological type T and its error. Col. (8): Distance in Mpc.
}
\end{deluxetable}

\begin{deluxetable}{rcccccccccccc}
\rotate
\tabletypesize{\scriptsize}
\tablecaption{UV Data in Our Catalog\label{tab2}}
\tablewidth{0pt}
\setlength{\tabcolsep}{0.04in}
\tablehead{
\colhead{PGC}&\multicolumn{2}{c}{Asymptotic Magnitudes}&\multicolumn{2}{c}{D25 Magnitudes}&
          \multicolumn{2}{c}{log $L$}&\multicolumn{2}{c}{Effective Radii}&
          \multicolumn{2}{c}{C31}&\multicolumn{2}{c}{C42}\\
\colhead{}&\multicolumn{1}{c}{FUV}&\multicolumn{1}{c}{NUV}&\multicolumn{1}{c}{FUV}&\multicolumn{1}{c}{NUV}&
          \multicolumn{1}{c}{FUV}&\multicolumn{1}{c}{NUV}&\multicolumn{1}{c}{FUV}&\multicolumn{1}{c}{NUV}&
          \colhead{FUV}&\colhead{NUV}&\colhead{FUV}&\colhead{NUV}\\
\colhead{}&\multicolumn{1}{c}{(mag)}&\multicolumn{1}{c}{(mag)}&\multicolumn{1}{c}{(mag)}&
\multicolumn{1}{c}{(mag)}&\multicolumn{1}{c}{(W)}&\multicolumn{1}{c}{(W)}&\multicolumn{1}{c}{(arcsec)}&
\multicolumn{1}{c}{(arcsec)}&\multicolumn{1}{c}{}&\multicolumn{1}{c}{}&
\multicolumn{1}{c}{}&\multicolumn{1}{c}{}\\
\multicolumn{1}{c}{(1)}&\multicolumn{1}{c}{(2)}&\multicolumn{1}{c}{(3)}&\multicolumn{1}{c}{(4)}&
\multicolumn{1}{c}{(5)}&\multicolumn{1}{c}{(6)}&\multicolumn{1}{c}{(7)}&\multicolumn{1}{c}{(8)}&
\multicolumn{1}{c}{(9)}&\multicolumn{1}{c}{(10)}&\multicolumn{1}{c}{(11)}&\multicolumn{1}{c}{(12)}&
\multicolumn{1}{c}{(13)}\\
}
\startdata
       12&   17.37 $\pm$ 0.01&   16.93 $\pm$ 0.03&   17.57 $\pm$ 0.01&   17.18 $\pm$ 0.01&  35.931&  35.929&   12.71&   13.19&    1.93&    2.12&    1.88&    2.10 \\
       62&          ...      &   18.48 $\pm$ 0.04&          ...      &   18.70 $\pm$ 0.04&     ...&  35.732&     ...&    8.67&     ...&    2.21&     ...&    2.15 \\
      120&   16.47 $\pm$ 0.04&   15.90 $\pm$ 0.05&   16.62 $\pm$ 0.01&   16.01 $\pm$ 0.01&  35.768&  35.816&   23.22&   22.42&    2.14&    2.14&    2.00&    2.02 \\
      129&   17.12 $\pm$ 0.03&   16.50 $\pm$ 0.02&   17.41 $\pm$ 0.02&   16.70 $\pm$ 0.01&  35.700&  35.768&   17.77&   15.89&    2.05&    2.11&    2.07&    2.03 \\
      143&   12.61 $\pm$ 0.04&   12.41 $\pm$ 0.03&   12.71 $\pm$ 0.00&   12.52 $\pm$ 0.00&  33.857&  33.759&   90.46&   91.78&    2.14&    2.16&    1.98&    2.05 \\
      176&          ...      &   16.29 $\pm$ 0.07&          ...      &   16.50 $\pm$ 0.01&     ...&  36.176&     ...&   14.76&     ...&    2.20&     ...&    2.12 \\
      192&          ...      &   17.12 $\pm$ 0.02&          ...      &   17.21 $\pm$ 0.01&     ...&  35.809&     ...&    8.53&     ...&    2.22&     ...&    2.19 \\
      215&          ...      &   16.83 $\pm$ 0.04&          ...      &   17.04 $\pm$ 0.02&     ...&  35.931&     ...&   13.52&     ...&    2.51&     ...&    2.49 \\
      243&   17.56 $\pm$ 0.05&   16.96 $\pm$ 0.02&   17.62 $\pm$ 0.01&   17.05 $\pm$ 0.01&  36.125&  36.187&   10.85&   10.92&    2.85&    3.18&    2.93&    3.14 \\
      250&   16.54 $\pm$ 0.08&   15.97 $\pm$ 0.02&   17.06 $\pm$ 0.01&   16.41 $\pm$ 0.01&  36.367&  36.417&   16.90&   15.49&    2.98&    2.94&    2.79&    2.88 \\
\enddata
\tablecomments{
Col.(1): PGC number. Col. (2)--(3): Asymptotic magnitude and its error in the FUV and NUV.
Cols. (4)--(5):  The aperture
magnitudes and their errors within the D25 ellipse in FUV and NUV, respectively.
Col. (6)--(7): Logarithm of the FUV and NUV luminosity. Col.
(8)--(9): Effective radii in the FUV and NUV. Col. (10)--(11): Concentration index C31 in the FUV and NUV.
Col. (12)--(13): Concentration index C42 in the FUV and NUV.
}
\end{deluxetable}

\begin{deluxetable}{rrrccrrrrrrr}
\rotate
\tabletypesize{\scriptsize}
\tablecaption{Achival Optical and Infrared Data in Our Catalog\label{tab3}}
\tablewidth{0pt}
\setlength{\tabcolsep}{0.04in}
\tablehead{
\colhead{PGC}&\multicolumn{4}{c}{Optical Magnitudes and Colors}&\multicolumn{3}{c}{2MASS Magnitudes}&
          \multicolumn{4}{c}{$IRAS$ Fluxes}\\
\colhead{}&\multicolumn{1}{c}{$B$}&\multicolumn{1}{c}{$I$}&\multicolumn{1}{c}{$(U-B)_{T}$}&\multicolumn{1}{c}{$(B-V)_{T}$}&
          \multicolumn{1}{c}{$J$}&\multicolumn{1}{c}{$H$}&\multicolumn{1}{c}{$K$}&\multicolumn{1}{c}{12$\mu$m}&
          \multicolumn{1}{c}{25$\mu$m}&\multicolumn{1}{c}{60$\mu$m}&\multicolumn{1}{c}{100$\mu$m}\\
\colhead{}&\multicolumn{1}{c}{(mag)}&\multicolumn{1}{c}{(mag)}&\multicolumn{1}{c}{(mag)}&
\multicolumn{1}{c}{(mag)}&\multicolumn{1}{c}{(mag)}&\multicolumn{1}{c}{(mag)}&\multicolumn{1}{c}{(mag)}&
\multicolumn{1}{c}{(Jy)}&\multicolumn{1}{c}{Jy}&\multicolumn{1}{c}{Jy}&
\multicolumn{1}{c}{Jy}\\

\multicolumn{1}{c}{(1)}&\multicolumn{1}{c}{(2)}&\multicolumn{1}{c}{(3)}&\multicolumn{1}{c}{(4)}&
\multicolumn{1}{c}{(5)}&\multicolumn{1}{c}{(6)}&\multicolumn{1}{c}{(7)}&\multicolumn{1}{c}{(8)}&
\multicolumn{1}{c}{(9)}&\multicolumn{1}{c}{(10)}&\multicolumn{1}{c}{(11)}&\multicolumn{1}{c}{(12)}\\
}
\startdata
       12& 14.66 $\pm$ 0.38&              ...&   ...&   ...& 12.10 $\pm$ 0.03& 11.38 $\pm$ 0.04& 11.11 $\pm$ 0.06&                       ...&                       ...&                       ...&                       ...\\
       62& 16.23 $\pm$ 0.21& 14.47 $\pm$ 0.08&   ...&   ...& 13.48 $\pm$ 0.08& 12.73 $\pm$ 0.09& 12.61 $\pm$ 0.14&                       ...&                       ...&                       ...&                       ...\\
      120& 13.01 $\pm$ 0.41& 11.29 $\pm$ 0.22&   ...&   ...& 10.33 $\pm$ 0.02&  9.59 $\pm$ 0.02&  9.33 $\pm$ 0.02&                       ...&                       ...&                       ...&                       ...\\
      129& 13.59 $\pm$ 0.30& 12.43 $\pm$ 0.22&   ...&   ...& 11.06 $\pm$ 0.02& 10.16 $\pm$ 0.02&  9.93 $\pm$ 0.03& $<$      0.38            & $<$      0.67            &          5.47 $\pm$  0.66&         14.29 $\pm$  1.86\\
      143& 10.89 $\pm$ 0.08&              ...&   ...&  0.40&              ...&              ...&              ...& $<$      0.12            & $<$      0.20            &          0.32 $\pm$  0.08&          1.04 $\pm$  0.26\\
      176& 14.33 $\pm$ 0.29& 12.77 $\pm$ 0.08&   ...&   ...& 11.86 $\pm$ 0.03& 11.24 $\pm$ 0.04& 10.82 $\pm$ 0.05& $<$      0.25            & $<$      0.33            &          1.21 $\pm$  0.12&          2.49 $\pm$  0.27\\
      192& 15.07 $\pm$ 0.36&              ...&   ...&   ...& 12.37 $\pm$ 0.03& 11.69 $\pm$ 0.04& 11.56 $\pm$ 0.07&                       ...&                       ...&                       ...&                       ...\\
      215& 15.49 $\pm$ 0.29& 14.38 $\pm$ 0.08&   ...&   ...& 14.09 $\pm$ 0.12& 13.23 $\pm$ 0.15& 12.99 $\pm$ 0.17&                       ...&                       ...&                       ...&                       ...\\
      243& 14.37 $\pm$ 0.36& 12.50 $\pm$ 0.32&   ...&  0.81& 11.25 $\pm$ 0.03& 10.63 $\pm$ 0.03& 10.30 $\pm$ 0.04&          0.11 $\pm$  0.03& $<$      0.21            &          0.37 $\pm$  0.05&          1.58 $\pm$  0.21\\
      250& 14.05 $\pm$ 0.54& 12.25 $\pm$ 0.10&   ...&   ...& 11.51 $\pm$ 0.04& 10.83 $\pm$ 0.05& 10.66 $\pm$ 0.06& $<$      0.25            & $<$      0.25            &          0.74 $\pm$  0.07&          3.64 $\pm$  0.66\\
      ...&              ...&              ...&   ...&   ...&              ...&              ...&              ...&                       ...&                  ...     &                  ...     &                ...       \\
\enddata
\tablecomments{
Col.(1): PGC number. Col. (2):  $B$-band total magnitude in Vega mag from HyperLeda.
Col. (3): $I$-band total magnitudes in Vega mag from HyperLeda.
Cols. (4)--(5):  Total asymptotic $(U-B)$ and $(B-V)$ colors from HyperLeda.
Cols. (6)--(8): 2MASS $J$, $H$, and $K$-band total magnitudes in Vega mag from \citep{Jarrett03} and the 2MASS Extended Source Catalog.
Cols. (9)--(12): IRAS 12, 25, 60, and 100$\mu$m fluxes in Jy from \citep{Rice88,Moshir90} and the $IRAS$ Point Source Catalog.
}
\end{deluxetable}

\end{document}